\documentclass[twocolumn,floatfix,pra,showpacs,preprintnumbers,superscriptaddress,amsmath,amssymb,10pt,aps,prl]{revtex4}

\usepackage{graphicx,subfigure,amsmath,bbm}

\usepackage{mathtools}
\usepackage{mathptmx}
\usepackage{subfigure}
\usepackage{psfrag,graphicx}
\usepackage{dcolumn}
\usepackage{amsmath,amssymb}
\usepackage{bm}
\usepackage{color}
\usepackage{latexsym}
\usepackage{epstopdf}
\usepackage{color}
\usepackage[english]{babel}
\usepackage{latexsym}
\usepackage{psfrag,graphicx}
\usepackage{subfigure}
\usepackage{amsmath}
\usepackage{amssymb}
\usepackage{amsfonts}
\usepackage{bm}
\usepackage{natbib}
\usepackage{epstopdf}
\usepackage{xcolor}
\usepackage{appendix}

\definecolor{mygrey}{gray}{0.35}
\definecolor{myblue}{rgb}{0.2,0.2,0.8}
\definecolor{myzard}{cmyk}{0,0,0.05,0}
\definecolor{mywhite}{rgb}{1,1,1}
\definecolor{mywhite}{rgb}{1,1,1}
\definecolor{myred}{rgb}{1,0.,0.3}

\usepackage[colorlinks=true,citecolor=myblue,linkcolor=myred]{hyperref}

\def\ba{\begin{align}}
\def\enda{\end{align}}
\def\bi{\begin{itemize}}
\def\ei{\end{itemize}}

\def\be{\begin{equation}}
\def\ee{\end{equation}}
\def\bea{\begin{eqnarray}}
\def\eea{\end{eqnarray}}
\def\bse{\begin{subequations}}
\def\ese{\end{subequations}}

\begin{document}
\title{Random Matrix Theory Approach to Quantum Fisher Information in Quantum Many-Body Systems}
\def\correspondingauthor{\footnote{Corresponding author: pivanov@phys.uni-sofia.bg}}
\author{Venelin P. Pavlov}
\affiliation{Center for Quantum Technologies, Department of Physics, St. Kliment Ohridski University of Sofia, James Bourchier 5 blvd, 1164 Sofia, Bulgaria}
\author{Yoana R. Chorbadzhiyska}
\affiliation{Center for Quantum Technologies, Department of Physics, St. Kliment Ohridski University of Sofia, James Bourchier 5 blvd, 1164 Sofia, Bulgaria}
\author{Charlie Nation}
\affiliation{Department of Physics and Astronomy, University College London, London WC1E 6BT, United Kingdom}
\author{Diego Porras}
\affiliation{Institute of Fundamental Physics IFF-CSIC, Calle Serrano 113b, 28006 Madrid, Spain}
\author{Peter A. Ivanov}
\affiliation{Center for Quantum Technologies, Department of Physics, St. Kliment Ohridski University of Sofia, James Bourchier 5 blvd, 1164 Sofia, Bulgaria}

\begin{abstract}
We theoretically investigate parameter quantum estimation in quantum chaotic systems. 
Our analysis is based on an effective description of non-integrable quantum systems in terms of a random matrix Hamiltonian.
Based on this approach we derive an analytical expression for the time evolution of the quantum Fisher information.
We test our random matrix theory prediction with the exact diagonalization of a non-integrable spin system, focusing on the estimation of a local magnetic field by measurements of the many-body state.
Our numerical calculations agree with the effective random matrix theory approach and show that the information on the local Hamiltonian parameter is distributed throughout the quantum system during the quantum thermalization process. 
Our analysis shows a first stage in which the initial information spread is quadratic in time which quickly passes into linear increase with slope determine by the decay rate of the measured spin observable. 
When the information is fully spread among all degrees of freedom a second quadratic time scale determines the long time behaviour of the quantum Fisher information.
\end{abstract}

\maketitle


\emph{Introduction.-} 
Initially excited quantum systems typically equilibrate to states exhibiting thermal properties, a process known as quantum thermalization \cite{Polkovnikov2011,Popescu2006,Rigol2008,Eisert2014,Alessio2016,Nandkishore2015}. 
At the core of our current understanding of this intriguing phenomenon is the Eigenstate Thermalization Hypothesis (ETH), which assumes that the many-body eigenstates of non-integrable Hamiltonians yield the same expectation values of local observables as those calculated with a microcanonical ensemble \cite{Deutsch1991,Deutsch2018,Srednicki1994,Srednicki1996}. 
The ETH can be formally expressed as a conjecture on the properties of matrix elements of local observables, which in turn can be derived from an effective description of non-integrable systems in terms of random matrix theory (RMT).
The validity of the ETH has been confirmed for a broad range of many-body systems by means of exact diagonalizations \cite{Rigol2009,Rigol2012,Swan2019,Jansen2019,Kim2014,Kirkova2023}. 
Furthermore, experimental quantum optical systems have allowed for investigation of quantum thermalization and the emergence of statistical physics in isolated quantum systems. Examples include recent experiments with ultracold atoms \cite{Gring2012,Kaufman2016}, trapped ions \cite{Smith2016,Clos2016,Kranzl2023}, and superconducting qubits \cite{Neill2016}. 

An important fundamental issue in quantum many-body theory is how information on local properties can be retrieved or estimated from observing the quantum system's dynamics. 
This problem is closely related to fundamental research on the connection between quantum chaos and scrambling of quantum information \cite{Swingle2018, Rampp2023} and also to applications like quantum metrology \cite{Qian2010,Song2012,Fiderer2018}. 
For example, the exponential sensitivity to small perturbation in imperfect time-reversal quantum dynamics is a widely studied signal for irreversibility \cite{Gorin2006,Schmitt2019,Kirkorova2022}. 
Here we address this question by investigating the dynamics of the quantum Fisher information (QFI) in quantum ergodic systems. 
The QFI is a quantity of central importance in quantum metrology. It quantifies the sensitivity of a given input state to a unitary transformation and provides the fundamental bound of the parameter estimation \cite{Paris2009,Pezze,Liu2020,Giovannetti2011}. 
The QFI also provides a sufficient condition to recognize entanglement in multiparticle state \cite{Pezze2009,Toth2012,Brenes2020,Pavlov2023}.

In this work we study the time-evolution of the QFI of non-integrable systems by a RMT approach. 
We model the Hamiltonian of the ergodic system as the sum of two contributions: a free non-interacting diagonal part and an interaction term modelled by a Gaussian orthogonal random matrix. This approach is valid as long as the coupling between a subsystem and the rest of a closed non-integrable system can be modelled as random matrix. Such an approach was originally proposed by J. Deutsch as a toy model to describe the emergence of quantum thermalization in isolated quantum systems \cite{Deutsch1991}. Recently, it was shown that this approach can be extended to predict the off-diagonal elements of observables, recovering the ETH \cite{Nation2018}. The description on RMT relies on strong assumptions that are basically equivalent to the ETH itself, however it allows to make scaling predictions that can be tested in experiments or exact numerical diagonalizations.

We test the results predicted from RMT by using an exact diagonalization of a non-integrable spin chain. The model consists of a system Hamiltonian describing one or a few non-interacting spins coupled with large spin system which plays the role of a finite quantum many-body bath. We find three time scales, which obey QFI. We show that in the beginning of the time evolution the QFI increases \emph{quadratically}. In this short time period the information of the parameter is not locally lost and the behaviour of the QFI resembles the standard quantum limit. After this time period the QFI quickly passes into a \emph{linear} increase with slope defined by the width of the random wave functions. Essentially, this width is the decay rate of the observable to the microcanonical average. In this second stage the information of the parameter propagates along the entire system. Remarkably, a second \emph{quadratic} time scale appears that determines the long-time behavior of QFI. It occurs when the information is spread among all quantum states involved in the evolution. We show that in this third stage the QFI is inversely proportional to the effective dimension of the system, a measure which quantifies the ergodicity of the system. The transition between the linear to quadratic time regimes occur at Heisenberg time determined by the density of states of the system.

For $N_{S}$ non-interacting spins one can expect that the QFI scales as $\sim N_{S}$ which gives the standard shot-noise limit. Crucially, for a few spins coupled to the quantum many-body heat bath, the system-bath interaction gives rise to a spin-spin correlation within the small subsystem. Hence, we show the quantum correlation may increase the QFI in a sense that it can exceeded the QFI corresponding to the standard quantum limit (SQL) without any initial entangled state preparation. 

\emph{Quantum parameter estimation.-} We consider a quantum system described by a non-integrable Hamiltonian, $\hat{H}(\lambda)=\hat{H}_{0}+\hat{H}_{I}$, consisting of non-interacting Hamiltonian $\hat{H}_{0}=\hat{H}_{\rm S}+\hat{H}_{\rm B}$ with $\hat{H}_{\rm S}$ and $\hat{H}_{\rm B}$ being the Hamiltonians for the subsystem and the many-body environment and respectively interaction part $\hat{H}_{I}$ describing the system-bath interaction. The eigenvectors and eigenenergies of the total Hamiltonian are 
$|\psi_{\mu}\rangle$ and $E_{\mu}$, such that $\hat{H}|\psi_{\mu}\rangle=E_{\mu}|\psi_{\mu}\rangle$. We also define non-interacting energy eigenbasis, $\hat{H}_{0}|\varphi_{\alpha}\rangle=E_{\alpha}|\varphi_{\alpha}\rangle$. The system is initially prepared in an out-of-equilibrium state $|\Psi_{0}\rangle=\sum_{\mu}a_{\mu}|\psi_{\mu}\rangle$ with $a_{\mu}=\langle\psi_{\mu}|\Psi_{0}\rangle$ which evolves under the action of the unitary propagator, $|\psi(\lambda)\rangle=e^{-i\hat{H}(\lambda)t}|\Psi_{0}\rangle$.
\begin{figure} 
\includegraphics[width=0.46\textwidth]{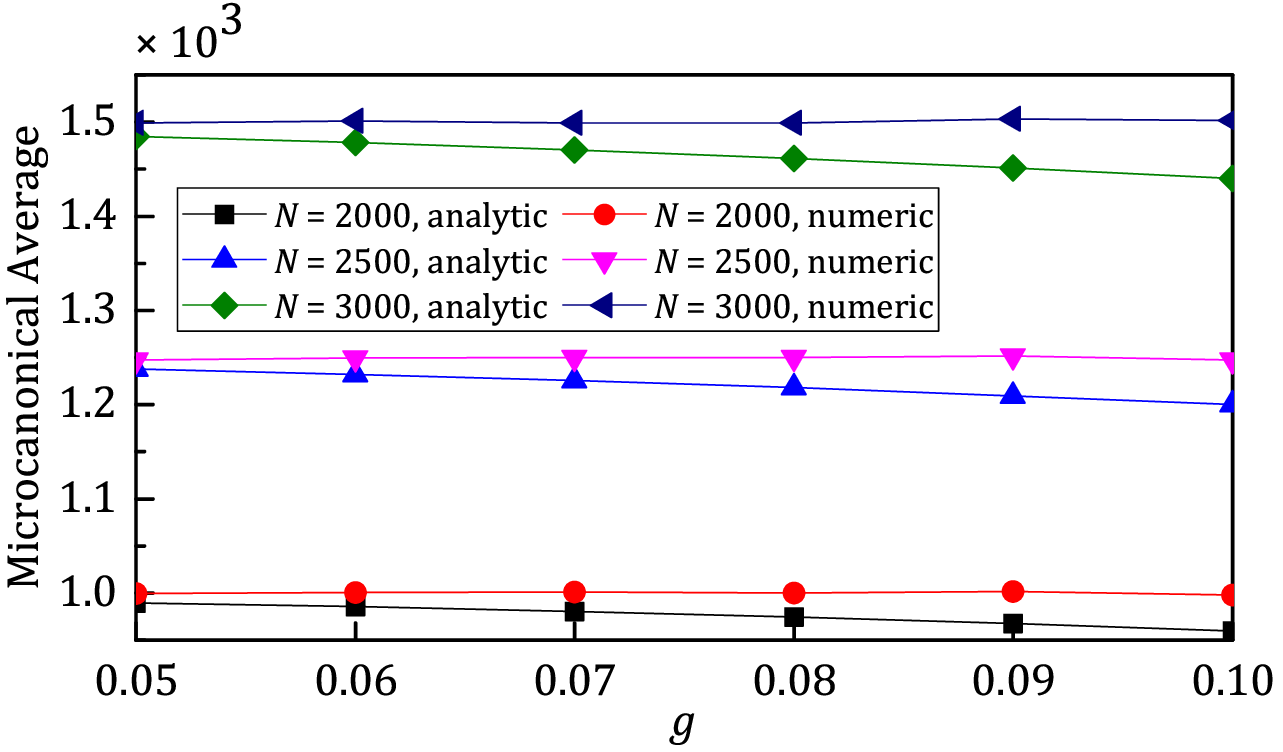}
\caption{Estimates for the microcanonical average for varying values of $g$ and $N$. Analytic results (\ref{mc_av}) are compared with the predictions given by the
diagonal ensemble (numerical results). Average over $10$ realizations of the random Hamiltonian is taken. The initial state is $|\Psi_{0}\rangle=|\varphi_{\alpha}\rangle$ with $\alpha=N/2$.}
\label{fig1}
\end{figure}

The classical Fisher information (CFI) $F_{\rm C}(\lambda)=\sum_{n}p(n|\lambda)^{-1}(\partial_{\lambda}p(n|\lambda))^{2}$
quantifies the amount of information on the parameter $\lambda$ which can be derived by performing discrete measurements with probability distribution $p(n|\lambda)$. The optimal strategy to measure the value of $\lambda$ is however associated with a privileged observable that maximizes the CFI. The CFI is upper bounded $F_{C}(\lambda)\leq F_{Q}(\lambda)$, where $F_{Q}(\lambda)$ is the QFI. The ultimate achievable precision of the parameter estimation is quantified via the quantum Cram\'er-Rao bound $\delta\lambda^{2}\geq 1/(MF_{Q}(\lambda))$ where $M$ is the repetition number.
The QFI $F_{Q}(\lambda)$ is a measure of distinguishability of the quantum states $|\psi(\lambda)\rangle$ and $|\psi(\lambda+d\lambda)\rangle$ with respect to the infinitesimal small variation of the parameter of interest $\lambda$. For pure state, the QFI reads, \cite{Paris2009}
\begin{equation}
F_{Q}(\lambda)=4(\langle\partial_{\lambda}\psi(\lambda)|\partial_{\lambda}\psi(\lambda)\rangle-|\langle\psi(\lambda)|\partial_{\lambda}\psi(\lambda)\rangle|^{2}).\label{QFI_def}
\end{equation}
The QFI can be expressed also by using the so-called symmetric logarithmic derivative (SLD) operator $\hat{L}_{\lambda}$ through the relation $F_{Q}(\lambda)=\langle\psi(\lambda)|\hat{L}_{\lambda}^{2}|\psi(\lambda)\rangle$. For pure state the SLD operator can be written as $\hat{L}_{\lambda}=2(|\partial_{\lambda}\psi(\lambda)\rangle\langle\psi(\lambda)|+|\psi(\lambda)\rangle\langle\partial_{\lambda}\psi(\lambda)|)$ and the optimal measurements that saturate the Cram\'er-Rao bound are projective measurements formed by the eigenvectors of $\hat{L}_{\lambda}$. 
\begin{figure}
\includegraphics[width=0.465\textwidth]{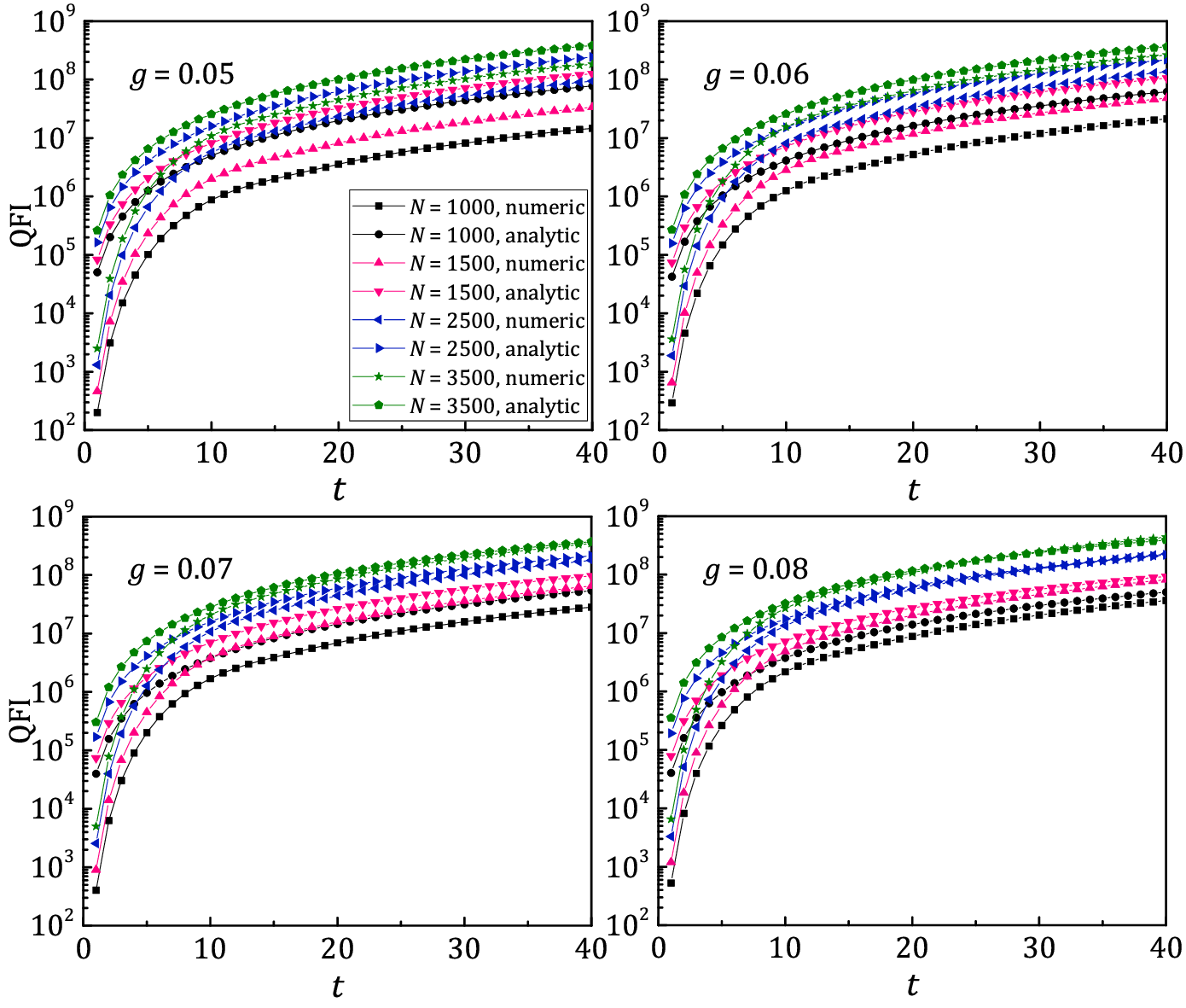}
\caption{Quantum Fisher information as a function of time for various $g$ and $N$. We compare the results for the QFI derived from the random matrix Hamiltonian using Eq. (\ref{QFI_def}) and the analytical result Eq. (\ref{main}). The random Hamiltonian is averaged over $10$ realizations. The initial state is $|\Psi_{0}\rangle=|\varphi_{\alpha}\rangle$ with $\alpha=N/2$ and $(\hat{H}^{\prime}_{0})_{\alpha\beta}=\alpha\delta_{\alpha\beta}$.}
\label{fig2}
\end{figure}

Hereafter we assume that the dependence on $\lambda$ comes only from the non-interacting Hamiltonian $\hat{H}_{0}(\lambda)$. This example corresponds to cases where $\lambda$ represents spin frequency or external magnetic field strength. Furthermore, we may express the QFI (\ref{QFI_def}) in the basis of $|\psi_{\mu}\rangle$ eigenvectors. We have \cite{sup}
\begin{eqnarray}
F_{Q}(\lambda)&=&4t^{2}\{\sum_{\mu\nu\rho}a^{*}_{\mu}a_{\nu}(\partial_{\lambda}\hat{H}_{0})_{\mu\rho}(\partial_{\lambda}\hat{H}_{0})_{\rho\nu}
e^{i\theta_{\mu\nu}t}{\rm sinc}(\theta_{\mu\rho}t)\notag\\
&&\times{\rm sinc}(\theta_{\rho\nu}t)\notag\\
&&-|\sum_{\mu\nu}a^{*}_{\mu}a_{\nu}e^{i\theta_{\mu\nu}t}(\partial_{\lambda}\hat{H}_{0})_{\mu\nu}{\rm sinc}(\theta_{\mu\nu}t)|^{2}\},\label{QFI_1}
\end{eqnarray}
where $(\partial_{\lambda}\hat{H}_{0})_{\mu\nu}=\langle\psi_{\mu}|\partial_{\lambda}\hat{H}_{0}|\psi_{\nu}\rangle$ are the matrix elements in the many-body interacting basis, $\theta_{\mu\nu}=(E_{\mu}-E_{\nu})/2$, and ${\rm sinc}(x)=\sin(x)/x$.
The expression (\ref{QFI_1}) is convenient for our further consideration because we can apply a RMT approach to evaluate the respective matrix elements.

\emph{Random Matrix Approach.-} Our analysis of the QFI is based on the random matrix model in which the non-interacting Hamiltonian $\hat{H}_{0}$ is modelled by diagonal matrix of size $N$, with $\omega$ being the constant spacing between the energy levels. The interaction term $\hat{H}_{I}$ is modelled by a random matrix, $(\hat{H}_{I})_{\alpha\beta}=h_{\alpha\beta}$, where $h_{\alpha\beta}$ are independent random numbers selected from the Gaussian orthogonal ensemble with probability distribution $P(h)\propto e^{-\frac{N}{4g^{2}}{\rm Tr} h^{2}}$, giving average $\langle h_{\alpha\beta}\rangle=0$, and variance $\langle h^{2}_{\alpha\beta}\rangle=g^{2}(1+\delta_{\alpha\beta})/N$ where $g$ is the coupling strength \cite{Deutsch1991,Nation2018}.

We expand the eigenstates of $\hat{H}$ in the non-interacting basis, $|\psi_{\mu}\rangle=\sum_{\alpha}c_{\mu}(\alpha)|\varphi_{\alpha}\rangle$, where $c_{\mu}(\alpha)$ are random variables whose statistical properties depend on the properties of the random matrix $\hat{H}_{I}$. The probability distribution of eigenstates is then described with Lorentzian, $\langle |c_{\mu}(\alpha)|^{2}\rangle_{V}=\Lambda(\mu,\alpha)=(\omega\Gamma/\pi)((E_{\mu}-E_{\alpha})^{2}+\Gamma^{2})^{-1}$ \cite{Deutsch1991, Nation2018},
where $\Gamma=\pi g^{2}/N\omega$ is the width of the wave function distribution and $\langle \ldots\rangle_{V}$ denotes an average over realizations of the random matrix $(\hat{H}_{I})_{\alpha\beta}$. The Lorentzian function is normalized such that $\sum_{\mu}\Lambda(\mu,\alpha)=\sum_{\alpha}\Lambda(\mu,\alpha)=1$. Furthermore, we assume a self-averaging condition where sum over random wave functions are replace with their ensemble average $\sum_{\alpha\ldots\beta}c_{\mu}(\alpha)\ldots c_{\nu}(\beta)=\sum_{\alpha\ldots\beta}\langle c_{\mu}(\alpha)\ldots c_{\nu}(\beta)\rangle_{V}$, see Supplemental Material (SM) \cite{sup}.

The self-averaging condition is essential for the evaluation of the QFI (\ref{QFI_1}), and is shown to hold in the description of observables in Ref. \cite{Nation2019_1, Dabelow2020}. Indeed, we can evaluate the sum of the matrix elements in (\ref{QFI_1}) in terms of averages of products of random-wave functions $c_{\mu}(\alpha)$. The treatment of $c_{\mu}(\alpha)$ as an independent random Gaussian variables, however, is not sufficient to determine the value of the off-diagonal matrix elements of an observable which are consistent with ETH \cite{Nation2018,Nation2019,Nation2019_1,Nation2020}. Because of that a non-Gaussian corrections should be included which arise as a result of the orthonormality condition. Based on a self-averaging condition the QFI (\ref{QFI_1}) is given by (see SM for more details \cite{sup})
\begin{eqnarray}
F_{Q}(\lambda)\approx4t^{2}\left\{\frac{\omega}{\pi\Gamma}(\hat{H}^{\prime2}_{0})_{\rm mc}
+\frac{(\Delta \hat{H}_{0}^{\prime2})_{\rm mc}}{2(\Gamma t)^{2}}
(e^{-2\Gamma t}-1+2\Gamma t)\right\}.\label{main}
\end{eqnarray}
This is the main result of our work. Here $\partial_{\lambda}\hat{H}_{0}=\hat{H}^{\prime}_{0}$ and $(\hat{H}^{\prime}_{0})_{\rm mc}$ is the microcanonical average of an observable $\hat{H}^{\prime}_{0}$ and $(\Delta \hat{H}_{0}^{\prime2})_{\rm mc}=(\hat{H}_{0}^{\prime2})_{\rm mc}-(\hat{H}^{\prime}_{0})^{2}_{\rm mc}$ is the microcanonical average of the variance of $\hat{H}^{\prime}_{0}$. The result (\ref{main}) is based on the following assumptions: (i) We assume \emph{sparsity} of $\hat{H}^{\prime}_{0}$ which implies that its matrix elements in the non-interacting basis is represented by a diagonal matrix or at least, by a matrix with only a few non-diagonal elements. (ii) We define \emph{smoothness} of the matrix elements of an observable in the following way: 
\begin{equation} \label{mc_av}
[(\hat{H}_{0}^{\prime})_{\alpha\alpha}]_{\mu}=\sum_{\alpha}\Lambda(\mu,\alpha)(\hat{H}^{\prime}_{0})_{\alpha\alpha},
\end{equation}
which represents the microcanonical average of the matrix elements $(\hat{H}^{\prime}_{0})_{\alpha\alpha}$ around the energy $E_{\mu}$, namely $(\hat{H}^{\prime}_{0})_{\rm mc}=[(\hat{H}_{0}^{\prime})_{\alpha\alpha}]_{\mu}$. Such an average is well defined as long as $\omega/\Gamma\ll 1$ which ensures that a large number of matrix elements are averaged in $[(\hat{H}_{0}^{\prime})_{\alpha\alpha}]_{\mu}$. Conditions for the validity of these assumptions have been discussed in detail in Ref. \cite{Nation_thesis}.

In Fig. \ref{fig1} we show the microcanonical average of observable $\hat{H}^{\prime}_{0}=\alpha\delta_{\alpha\beta}$ according to (\ref{mc_av}) compared with the diagonal average $\langle \bar{H}^{\prime}_{0}\rangle ={\rm Tr}(\hat{H}^{\prime}_{0}\hat{\rho}_{\rm DE})$, where $\hat{\rho}_{\rm DE}=\sum_{\mu}|a_{\mu}|^{2}|\psi_{\mu}\rangle\langle\psi_{\mu}|$ is the density matrix of the diagonal ensemble.
In Fig. \ref{fig2} we show the time evolution of the QFI where we set $\lambda=\omega$. We compare the exact result based on Eq. (\ref{QFI_def}) using the random matrix model and analytical expression (\ref{main}). The numerical and analytical results are averaged over $10$ realizations of the random Hamiltonian. We set the initial state to be an eigenstate of the non-interacting Hamiltonian, $|\Psi_{0}\rangle=|\varphi_{\alpha}\rangle$ with $\alpha$ selected at the middle of the energy spectrum. We compare the results for various $g$ and $N$. We see a good agreement between both results except for the initial time evolution. We observe however that the relative error between both results decreases as we increase the matrix size $N$ \cite{sup}. We also test the result for other initial states $|\varphi_{\alpha}\rangle$. We find that the relative error convergences faster for initial states at the lower half of the non-interacting energy spectrum and respectively slower at the upper half.

\emph{Exact Diagonalization.-} We now turn to the comparison of our main result (\ref{main}) with exact diagonalization of non-integrable spin chain. We consider 1D spin system with a Hamiltonian of the form
\begin{equation}
\hat{H}=\hat{H}_{\rm S}+\hat{H}_{\rm B}+\hat{H}_{\rm SB}.\label{spinH}
\end{equation}
The system Hamiltonian describes a single spin in a presence of a $B$-field
\begin{equation}
\hat{H}_{\rm S}=B\sigma^{z}_{1},
\end{equation}
where $\sigma^{q}_{j}$ ($q=x,y,z$) are the Pauli matrices acting on $j$-th site and $B$ is the parameter we wish to estimate, namely $\lambda=B$. The bath Hamiltonian describes a spin chain with Ising interaction
\begin{equation}
\hat{H}_{\rm B}=\sum_{k>1}^{N} B_x^{(\rm B)}\sigma_k^{x} + \sum_{k>1}^{N-1}J_x(\sigma^{+}_{k}\sigma^{-}_{k+1}+\sigma^{-}_{k}\sigma^{+}_{k+1}),
\end{equation}
where $B^{(B)}_{x}$ is the magnetic field along the $x$-axis and $J_x>0$
is the spin-spin coupling. The interaction Hamiltonian describes a coupling between the system spin and a single bath spin of index $r$ 
\begin{equation}
\hat{H}_{\rm SB}=J_z^{(\rm SB)}\sigma^{z}_{1}\sigma^{z}_{r}+J_x^{(\rm SB)}(\sigma^{+}_{1}\sigma^{-}_{r}+\sigma^{-}_{1}\sigma^{+}_{r}),
\end{equation}
with coupling strengths $J^{(\rm SB)}_{z}$ and $J^{(\rm SB)}_{x}$.

As long as the non-integrable system is well described by RMT we expect that expression Eq. (\ref{main}) holds with the modification $\omega\rightarrow 1/D(E_{0})$ where $D(E_{0})$ is the density of states at the initial energy $E_{0}$ \cite{Nation2018}. This assumption holds when the energy scale over which the density of states changes is large with respect to $\Gamma$, the energy width of the random wave functions. This limits the above approach to a weak coupling regime.  

Then using Eq. (\ref{main}) with $\hat{H}_{0}^{\prime}=\sigma^{z}_{1}$ and thereby $(\hat{H}^{\prime 2}_{0})_{\rm mc}=1$ and $(\Delta \hat{H}_{0}^{\prime2})_{\rm mc}=1-(\sigma^{z}_{1})^{2}_{\rm mc}$, the QFI becomes
\begin{equation}\label{mainSpin}
F_{Q}(B)\approx 4t^{2}\left\{\frac{1}{\pi D(E_{0})\Gamma}+\frac{1-(\sigma_{1}^{z})^{2}_{\rm mc}}{2(\Gamma t)^{2}}
(e^{-2\Gamma t}-1+2\Gamma t)\right\}.
\end{equation}
Note that in applying the QFI as above we are describing a local observable of the spin system in terms of RMT. In the SM \cite{sup} we show that such local observables are indeed well described in terms of RMT as long as certain energy scales of the system are sufficiently separated. Notably, the sparsity condition above follows trivially for a local observable \cite{Nation_thesis}.

In order to find the value of $\Gamma$ we use that the time dependence of an observable $\hat{O}$ obeys \cite{Nation2019,Dabelow2020}
\begin{equation}
\langle \hat{O}(t)\rangle =\langle \hat{O}(t)\rangle_{0}e^{-2\Gamma t}+\langle \bar{O}\rangle(1-e^{-2\Gamma t})\label{O},
\end{equation}
where $\langle \hat{O}(t)\rangle_{0}$ is the evolution of the observable according the non-interaction Hamiltonian $\hat{H}_{0}$ and $\langle \bar{O}\rangle$ is the time-average value defined by $\langle \bar{O}\rangle={\rm Tr}(\hat{O}\hat{\rho}_{\rm DE})$. Thermalization of a closed system implies the equality $\langle \bar{O}\rangle\approx(\hat{O})_{\rm mc}$.

\begin{figure}
\includegraphics[width=0.45\textwidth]{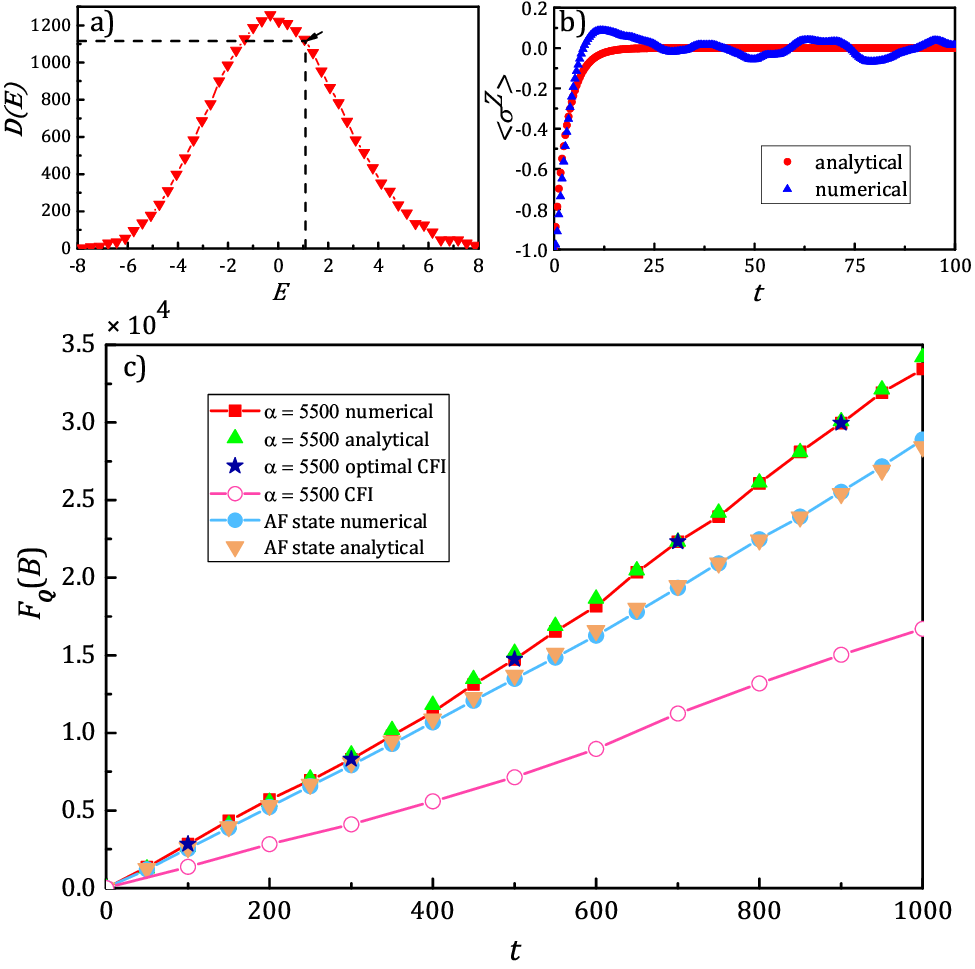}
\caption{(a) Density of states, as a function of energy for $N=13$ and initial state $|\Psi_0\rangle = |\varphi_\alpha \rangle$ with $\alpha=5500$. The dashed lines show the point $D(E_0)$, where $E_0 = \langle \Psi_{0} |\hat{H}| \Psi_{0} \rangle$. (b) Time evolution of the spin system observable $\sigma^{z}_{1}$ both numerically and analytically using Eq. (\ref{O}). (c) Quantum Fisher information as a function of time for initial state $|\Psi_0\rangle = |\varphi_\alpha \rangle$ with $\alpha=5500$ for $N=13$ and for antiferromagnetic initial state $|\Psi_0\rangle =|\uparrow \rangle_S |\downarrow\uparrow \downarrow ... \rangle_B$ for $N=15$. We compare the exact result for the QFI (\ref{QFI_def}) with Hamiltonian (\ref{spinH}) and the analytical expression (\ref{mainSpin}). The parameters are set to $B=0.01$, $B_x^{(\rm B)}=0.3$, $J_z^{(\rm SB)}=0.2$, $ J_x^{(\rm SB)}=0.4$, $J_x =1$, and $r = 5$. The decay rate is $\Gamma=0.15$ for both $N=13$ and $N=15$ is calculated by using Eq (\ref{O}) to fit the exact time evolution of the $\sigma^{z}_{1}$ operator for the system spin. The density of states $D(E_0)$ is evaluated by interpolation of $D(E)$.}
\label{fig3}
\end{figure}

In Fig. \ref{fig3}(a) we plot the density of states as a function of the energy. From here we can extract the value of $D(E_{0})$ at the initial energy $E_{0}$. We also fit the time evolution of observable $\sigma^{z}_{1}$ to obtain the value of $\Gamma$, see inset of Fig. \ref{fig3}(b). In Fig. \ref{fig3}(c) we show the comparison between the exact result for the QFI using (\ref{QFI_def}) and the analytical expression (\ref{mainSpin}) for various initial states. Since the small spin sub-system thermalizes the information for the parameter $B$ is locally lost. What we see, however, is that because of the spin-spin interaction the information has been not lost, but spread among the other degrees of freedom. At the beginning of the time evolution for $t \lesssim (2\Gamma)^{-1}$ the information flow is quadratic, where $F_{Q}(B)\approx 4t^{2} (\Delta \hat{H}_{0}^{\prime2})_{\rm mc}$. In this first stage the information for the parameter is still not locally lost and it can be determined with uncertainty bounded by the SQL. After this time period the growth of QFI becomes linear in time with slope determined by the decay rate $\Gamma$, namely $F_{Q}(B)\approx (4 t/\Gamma)(\Delta \hat{H}_{0}^{\prime2})_{\rm mc}$. In this second stage the information flow propagates along the entire system. Remarkably, a second quadratic time scale defines the long time behaviour of the QFI, that occurs when the information has fully spread between all degrees of freedom. In this third case we have $F_{Q}(B)\approx (4 t^{2}/\pi D(E_{0})\Gamma)(\hat{H}^{\prime2}_{0})_{\rm mc}$. In fact, we may connect the third time scaling of QFI with the effective dimension of the system. The effective dimension is defined by $d_{\rm eff}=(\sum_{\mu}|\langle\Psi(0)|\psi_{\mu}\rangle|^{4})^{-1}$ and provides an estimation for the ergodicity of a system. Also the mean amplitude of time fluctuations of an observable are bounded by $d_{\rm eff}^{-1/2}$ \cite{Linden2009}. The condition $d_{\rm eff}\gg 1$ implies that the initial state is composed of a large number of energy eigenstates which leads to suppression of temporal fluctuations of an observable and equilibration of the system. Using RMT approach it can be shown that $d_{\rm eff}=(2\pi/3) D(E_{0})\Gamma$ \cite{Nation2018}. Hence the long-time behaviour of the QFI becomes $F_{Q}(B)\approx (8t^2 /3 d_{\rm eff})(\hat{H}^{\prime2}_{0})_{\rm mc}$. This relation indicates that the final quadratic behaviour of the QFI occurs when the information of the parameter has been distributed over all quantum states involved in the evolution of the quantum system. The crossover between the linear to quadratic time regimes occurs at the Heisenberg time $\tau\approx \pi D(E_{0}) ((\Delta \hat{H}_{0}^{\prime2})_{\rm mc}/(\hat{H}^{\prime2}_{0})_{\rm mc})$, which is longest time scale for the system \cite{Schiulaz2019}. We point out that the density of states is related with the microcanonical entropy $S$ via the relation $1/D(E_{0})=e^{-S}$. Since the entropy is extensive quantity, the transition time $\tau$ increases with the number of spins.

\begin{figure}
\includegraphics[width=0.45\textwidth]{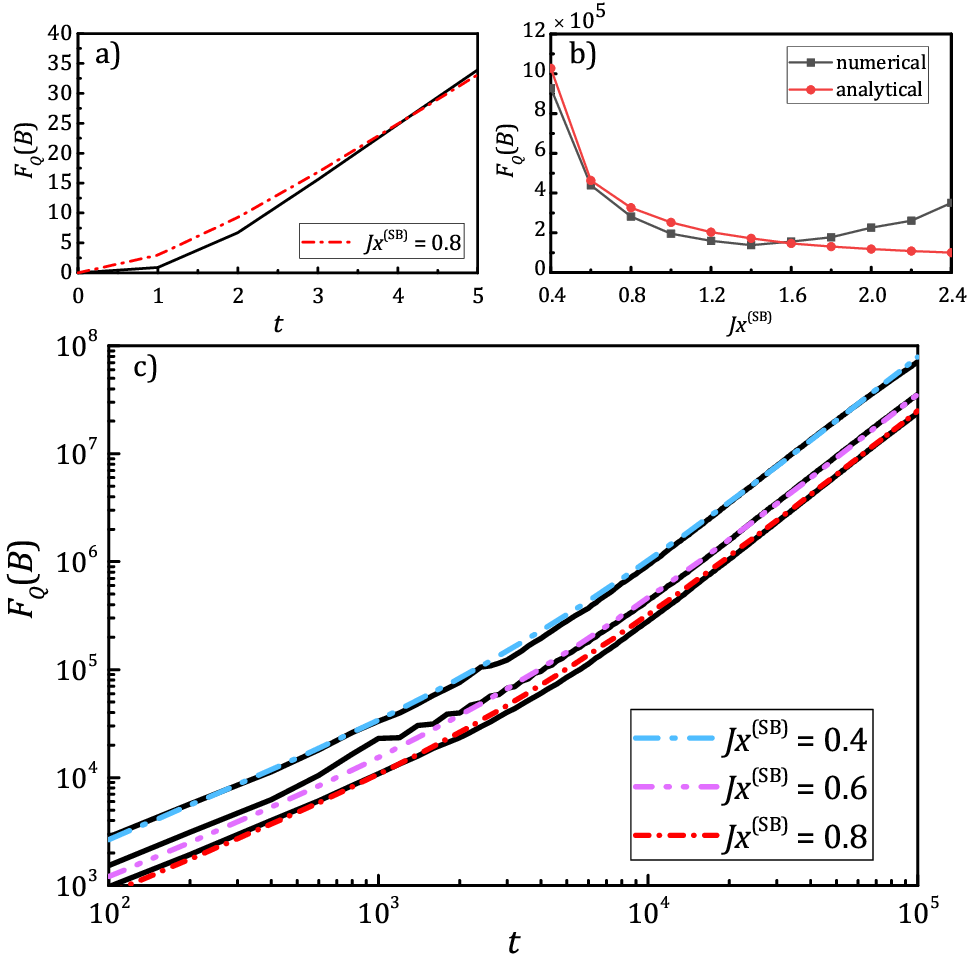}
\caption{a) Short time evolution of QFI for chain with $N=13$ spins. b) The QFI for various $J^{\rm SB}_{x}$ for $t=10^{4}$. c) Long time evolution of the QFI. The initial state is $|\Psi_{0}\rangle=|\varphi_{\alpha}\rangle$ for $\alpha=5500$.  We compare the numerical results (solid lines) with the analytical expression (\ref{mainSpin}).}
\label{fig5}
\end{figure}

In Fig. \ref{fig5} we plot the short and long time behaviour of the QFI. We observe a good agreement between the exact and the analytical results. As we see increasing the time the QFI makes a transition to quadratic time regime. Increasing the spin-bath coupling $J^{(\rm SB)}_{x}$ leads to higher decay rate $\Gamma$ which lower the QFI according to Eq (\ref{mainSpin}). We note that for larger spin-bath couplings, we expect the same general phenomena, however with differing functional forms of the random wave function distribution. For example, at intermediate couplings it has been observed that the random wave function takes a Gaussian form \cite{Atlas2017}, and for strong couplings where a full random matrix Hamiltonian is valid, the density of states dominates the energy dependence and hence leads to decay in the form of a Bessel function \cite{Herrera2014, Herrera2018}. In each case a RMT approach holds, however the assumption here of Lorentzian wave functions is strictly valid for weak couplings.

An important issue is whether we can recover the behaviour of the QFI by measuring a suitable observable. An optimal measurement that provides equality between CFI and QFI is given by the eigenvectors of the SLD operator $\hat{L}_{B}$. We numerically diagonalize $\hat{L}_{B}$ and respectively calculate the CFI as is shown in Fig. \ref{fig3}. Such a basis, however, is composed by an entangled states and is not suitable for measurement. A more convenient approach is to detect the spin populations $p_{s_{1},\ldots,s_{N}}={\rm Tr}(\hat{\rho}(t)\hat{\Pi}_{s_{1},\dots,s_{N}})$, where $\hat{\rho}(t)=|\psi(t)\rangle\langle\psi(t)|$ is the density operator and $\hat{\Pi}_{s_{1},\dots,s_{N}}$ is the projection operator with $s_{l}=\uparrow_{l},\downarrow_{l}$. In Fig. (\ref{fig3}) we plot the CFI. We see that although the CFI is lower than QFI the main properties of the QFI are captured by detection the spin populations.  

We proceed with an application of our result (\ref{mainSpin}) to a spin system Hamiltonian consisting of two spins, $\hat{H}_{\rm S}=B(\sigma^{z}_{1}+\sigma^{z}_{2})$. In that case we have $(\hat{H}^{\prime 2}_{0})_{\rm mc}=(\Delta \hat{H}_{0}^{\prime2})_{\rm mc}=2+2(\sigma^{z}_{1}\sigma^{z}_{2})_{\rm mc}$. We see that as long as there is no correlation between the system spins, $(\sigma^{z}_{1}\sigma^{z}_{2})_{\rm mc}=(\sigma^{z}_{1})_{\rm mc}(\sigma^{z}_{2})_{\rm mc}\approx 0$, the QFI is twice the QFI for a single system spin, $F^{(\rm SQL)}_{Q}(B)=2 F_{Q}(B)$, which corresponds to the SQL. Remarkably, we find a special system-bath interaction which leads to correlation between the two spins in a sense that $(\sigma^{z}_{1}\sigma^{z}_{2})_{\rm mc}> 0$. Therefore, the corresponding QFI becomes $F^{(1)}_{Q}(B)>F^{(\rm SQL)}_{Q}(B)$ and thus one can overcome the SQL without initial entangled state preparation. We give additional details on this application in the SM \cite{sup}.

\emph{Conclusion.-} We derive a random matrix approach to the time evolution of the QFI in a quantum ergodic system. The QFI is a measure for distinguishability of two quantum states with respect to infinitesimal variation of some parameter. Our analysis is based on a random Hamiltonian that yields an approximate description of a quantum non-integrable system. We show that the initial time growth of the QFI is quadratic, which quickly passes into a linear with slope defined by the width of the random wave function. Furthermore, we have found a second quadratic time scale which determines the long time behaviour of the QFI. This timescale is shown to correspond to the Heisenberg time, after which information of the local observable has spread throughout all accessible degrees of freedom of the system.


We compared our RMT result with exact diagonalization of non-integrable spin chain, confirming the RMT prediction of three separate timescales. We have shown that the information for a parameter describing single spin system is locally lost but propagates among the other degrees of freedom of the spin system. The transition time between the linear and quadratic time regimes depends on the density of states, and increases with the number of spins.

\emph{Acknowledgments.-} V. P. P. and P. A. I. acknowledge the Bulgarian national plan for recovery and resilience, contract BG-RRP-2.004-0008-C01 (SUMMIT: Sofia University Marking Momentum for Innovation and Technological Transfer), project number 3.1.4. C. N. acknowledges support from the Engineering and Physical Sciences Research Council (EPSRC UK) and the Gordon and Betty Moore Foundation grant GBMF8820. D. P. acknowledges support from the Spanish project PID2021-127968NB-I00 funded by MCIN/AEI/10.13039/501100011033/FEDER, UE.


\clearpage
\pagebreak
\widetext
\begin{center}
\textbf{\large Supplemental Material: Random Matrix Theory Approach to Quantum Fisher Information in Quantum Many-Body Systems 
}
\end{center}
\begin{center}
Venelin P. Pavlov$^1$, Yoana R. Chorbadzhiyska$^1$, Charlie Nation$^2$, Diego Porras$^3$, and Peter A. Ivanov$^1$ 
\end{center}
\begin{center}
$^1$\emph{Center for Quantum Technologies, Department of Physics, St. Kliment Ohridski University of Sofia, James Bourchier 5 blvd, 1164 Sofia, Bulgaria}\\
$^2$\emph{Department of Physics and Astronomy, University College London, London WC1E 6BT, United Kingdom}\\
$^3$\emph{Institute of Fundamental Physics IFF-CSIC, Calle Serrano 113b, 28006 Madrid, Spain}
\end{center}

\setcounter{equation}{0}
\setcounter{figure}{0}
\setcounter{table}{0}
\setcounter{page}{1}
\makeatletter
\renewcommand{\theequation}{S\arabic{equation}}
\renewcommand{\thefigure}{S\arabic{figure}}
\renewcommand{\bibnumfmt}[1]{[S#1]}
\renewcommand{\citenumfont}[1]{S#1}


\begin{section}{Quantum Fisher Information in the many-body interacting basis}\label{QFI}

The quantum Fisher information for a pure state is given by
\begin{equation}
F_{Q}(\lambda)=4\{\langle\partial_{\lambda}\psi|\partial_{\lambda}\psi\rangle-\langle\psi|\partial_{\lambda}\psi\rangle\langle\partial_{\lambda}\psi|\psi\rangle\},\label{Fisher}
\end{equation}
where the state vector is $|\psi\rangle=e^{-i\hat{H} t}|\Psi_{0}\rangle$ and $|\Psi_{0}\rangle$ is the initial state which is independent on parameter $\lambda$. Therefore, we have $|\partial_{\lambda}\psi\rangle=(\partial_{\lambda}e^{-i\hat{H} t})|\Psi_{0}\rangle$. The partial derivative can be written as
\begin{equation}
\partial_{\lambda}e^{-i\hat{H} t}=-it\int_{0}^{1}ds e^{-i\hat{H} t}e^{i\hat{H} t s}(\partial_{\lambda}\hat{H})e^{-i\hat{H} t s}.\label{D}
\end{equation}
The quantum Fisher information can be rewritten as
\begin{equation}
F_{Q}(\lambda)=4\{\langle\Psi_{0}|(\partial_{\lambda}e^{i\hat{H} t})(\partial_{\lambda}e^{-i\hat{H} t})|\Psi_{0}\rangle
-|\langle\Psi_{0}|e^{i\hat{H} t}(\partial_{\lambda}e^{-i\hat{H} t})|\Psi_{0}\rangle|^{2}\}.
\end{equation}
Now, let's consider the first term. We have
\begin{equation}
\langle\Psi_{0}|(\partial_{\lambda}e^{i\hat{H} t})(\partial_{\lambda}e^{-i\hat{H} t})|\Psi_{0}\rangle=
\sum_{\mu\nu\rho}a^{*}_{\mu}a_{\nu}\langle\psi_{\mu}|\partial_{\lambda}e^{i\hat{H}t}e^{-i\hat{H}t}|\psi_{\rho}\rangle\langle\psi_{\rho}|
e^{i\hat{H}t}\partial_{\lambda}e^{-i\hat{H}t}|\psi_{\nu}\rangle,\label{11}
\end{equation}
where we use that $|\Psi_0\rangle = \sum_\mu a_\mu |\psi_\mu\rangle$ and $\sum_{\rho}|\psi_{\rho}\rangle\langle\psi_{\rho}|=\textbf{1}$. Using (\ref{D}) we obtain
\begin{eqnarray}
\langle\psi_{\rho}|e^{i\hat{H}t}\partial_{\lambda}e^{-i\hat{H}t}|\psi_{\nu}\rangle&=&
-it\int_{0}^{1}ds\langle\psi_{\rho}|e^{is\hat{H}t}\partial_{\lambda}\hat{H}e^{-is\hat{H}t}|\psi_{\nu}\rangle
=-i t\langle\psi_{\rho}|\partial_{\lambda}\hat{H}|\psi_{\nu}\rangle\int_{0}^{1}ds e^{i(E_{\rho}-E_{\nu})st}\notag\\
&&=-it\langle\psi_{\rho}|\partial_{\lambda}\hat{H}|\psi_{\nu}\rangle e^{i\theta_{\rho\nu}t}{\rm sinc}(\theta_{\rho\nu}t),
\end{eqnarray}
Here we have defined $\theta_{\mu\nu}= \frac{E_\mu - E_\nu}{2}$. Therefore, we get
\begin{equation}
\langle\Psi_{0}|(\partial_{\lambda}e^{i\hat{H} t})(\partial_{\lambda}e^{-i\hat{H} t})|\Psi_{0}\rangle=t^{2}
\sum_{\mu\nu\rho}a^{*}_{\mu}a_{\nu}\langle\psi_{\mu}|\partial_{\lambda}\hat{H}|\psi_{\rho}\rangle\langle\psi_{\rho}|\partial_{\lambda}\hat{H}|\psi_{\nu}\rangle e^{i\theta_{\mu\nu}}
{\rm sinc}(\theta_{\mu\rho}t){\rm sinc}(\theta_{\rho\nu}t).\label{FT}
\end{equation}

Similarly for the second term we obtain
\begin{eqnarray}
\langle\Psi_{0}|e^{i\hat{H} t}(\partial_{\lambda}e^{-i\hat{H} t})|\Psi_{0}\rangle
=-it\sum_{\mu\nu}a^{*}_{\mu}a_{\nu}\int_{0}^{1}ds\langle\psi_{\mu}|e^{is\hat{H}t}\partial_{\lambda}\hat{H}e^{-is\hat{H}t}|\psi_{\nu}\rangle
=-it\sum_{\mu\nu}a^{*}_{\mu}a_{\nu}e^{i\theta_{\mu\nu}t}\langle\psi_{\mu}|\partial_{\lambda}\hat{H}|\psi_{\nu}\rangle{\rm sinc}(\theta_{\mu\nu}t).\label{22}
\end{eqnarray}
Using (\ref{Fisher}), (\ref{FT}) and (\ref{22}) we obtain

\begin{eqnarray}
F_{Q}(\lambda)=4t^{2}\left\{\sum_{\mu\nu\rho}a^{*}_{\mu}a_{\nu}(\partial_{\lambda}\hat{H}_{0})_{\mu\rho}(\partial_{\lambda}\hat{H}_{0})_{\rho\nu}
e^{i\theta_{\mu\nu}t}{\rm sinc}(\theta_{\mu\rho}t)
{\rm sinc}(\theta_{\rho\nu}t)
-|\sum_{\mu\nu}a^{*}_{\mu}a_{\nu}e^{i\theta_{\mu\nu}t}(\partial_{\lambda}\hat{H}_{0})_{\mu\nu}{\rm sinc}(\theta_{\mu\nu}t)|^{2}\right\}.
\end{eqnarray}

\end{section}

\begin{section}{Correlation Functions}\label{CF}
In this section we outline the core RMT approach to eigenstate correlations formulated in Ref. \cite{a}. We can calculate arbitrary correlation functions of the random wave functions $c_{\mu}(\alpha)$ by defining the respective generating function. For $\mu=\nu$ it reads
\begin{equation}
G_{\mu\mu}(\vec{\xi}_{\mu})\propto e^{\frac{1}{2}\sum_{\alpha}\xi^{2}_{\mu,\alpha}\Lambda(\mu,\alpha)}.\label{CF11}
\end{equation}
where $\vec{\xi}_{\mu}=(\xi_{\mu,1},\xi_{\mu,2}\ldots \xi_{\mu,N})$ are ancillary fields. An arbitrary correlation function of the random wave functions can be obtained via 
\begin{equation}
\langle c_{\mu}(\alpha)c_{\mu}(\alpha^{\prime})\ldots c_{\mu}(\beta)c_{\mu}(\beta^{\prime})\rangle_{V}=\frac{1}{G_{\mu\mu}}\partial_{\xi_{\mu,\alpha}}\partial_{\xi_{\mu,\alpha^{\prime}}}\ldots \partial_{\xi_{\mu,\beta}}\partial_{\xi_{\mu,\beta^{\prime}}} \left. G_{\mu\mu}\right|_{\xi_{\mu,\alpha}=0}.
\end{equation}
Similarly for $\mu\neq\nu$ the generating function is
\begin{equation}
G_{\mu\nu}(\vec{\xi}_{\mu},\vec{\xi}_{\nu})\propto e^{\frac{1}{2}\sum_{\alpha}\xi^{2}_{\mu,\alpha}\Lambda(\mu,\alpha)+\frac{1}{2}\sum_{\alpha}\xi^{2}_{\nu,\alpha}\Lambda(\nu,\alpha)-\frac{1}{2}\sum_{\alpha\beta}
\xi_{\mu,\alpha}\xi_{\mu,\beta}\xi_{\nu,\alpha}\xi_{\nu,\beta}\frac{\Lambda(\mu,\alpha)\Lambda(\mu,\beta)\Lambda(\nu,\alpha)\Lambda(\nu,\beta)}{\sum_{\gamma}\Lambda(\mu,\gamma)
\Lambda(\nu,\gamma)}}.\label{GF2}
\end{equation}
and the correlation function becomes
\begin{equation}
\langle c_{\mu}(\alpha)c_{\nu}(\alpha^{\prime})\ldots c_{\mu}(\beta)c_{\nu}(\beta^{\prime})\rangle_{V}=\frac{1}{G_{\mu\nu}}\partial_{\xi_{\mu,\alpha}}\partial_{\xi_{\nu,\alpha^{\prime}}}\ldots \partial_{\xi_{\mu,\beta}}\partial_{\xi_{\nu,\beta^{\prime}}} \left. G_{\mu\nu}\right|_{\xi_{\mu,\alpha}=0,\xi_{\nu,\alpha}=0}.
\end{equation}

In order to evaluate the QFI we focus on two sets of four-point correlation functions of interest: $\langle c_{\mu}(\alpha)c_{\nu}(\beta)c_{\mu}(\alpha^{\prime})c_{\nu}(\beta^{\prime})\rangle_{V}$ for $\mu=\nu$ and $\mu\neq\nu$. 

For $\mu=\nu$ the random wave functions can be treated as an independent random variables, namely
\begin{equation}
\langle c_{\mu}(\alpha)c_{\mu}(\beta)c_{\mu}(\alpha^{\prime})c_{\mu}(\beta^{\prime})\rangle_{V}=
\Lambda(\mu,\alpha)\Lambda(\mu,\beta)\delta_{\alpha\alpha^{\prime}}\delta_{\beta\beta^{\prime}}+\Lambda(\mu,\alpha)
\Lambda(\mu,\alpha^{\prime})(\delta_{\alpha^{\prime}\beta^{\prime}}\delta_{\alpha\beta}+\delta_{\alpha\beta^{\prime}}\delta_{\alpha^{\prime}\beta}).
\end{equation}
For $\mu\neq\nu$ we have
\begin{equation}
\langle c_{\mu}(\alpha)c_{\nu}(\beta)c_{\mu}(\alpha^{\prime})c_{\nu}(\beta^{\prime})\rangle_{V}=\Lambda(\mu,\alpha)\Lambda(\nu,\beta)\delta_{\alpha\alpha^{\prime}}\delta_{\beta\beta^{\prime}}
-\frac{\Lambda(\mu,\alpha)\Lambda(\nu,\beta)\Lambda(\mu,\alpha^{\prime})\Lambda(\nu,\beta^{\prime})}{\sum_{\gamma}\Lambda(\mu,\gamma)\Lambda(\nu,\gamma)}(\delta_{\alpha\beta}\delta_{\alpha^{\prime}\beta^{\prime}}+\delta_{\alpha\beta^{\prime}}\delta_{\alpha^{\prime}\beta}).\label{NG}
\end{equation}
The first term in (\ref{NG}) describes the four-point correlation function as an independent random Gaussian variables, while the last two terms correspond to the non-Gaussian correction which arises as a result of the orthogonality condition.

In fact we may express graphically the correlation functions as a sum of products of two-point correlation functions. Consider first $\mu=\nu$. Then we have  
\begin{equation}
\langle c_{\mu}(\alpha)c_{\mu}(\beta)c_{\mu}(\alpha^{\prime})c_{\mu}(\beta^{\prime})\rangle_{V}=
\langle c_{\mu}(\overbracket{\alpha)c_{\mu}(\beta})c_{\mu}(\overbracket{\alpha^{\prime})c_{\mu}(\beta^{\prime}})\rangle_{V}
+\langle c_{\mu}(\overbracket{\alpha)c_{\mu}(\underbracket{\beta)c_{\mu}(\alpha^{\prime}})c_{\mu}(\beta^{\prime}})\rangle_{V}
+\langle c_{\mu}(\lefteqn{\overbracket{\phantom{\alpha)c_{\mu}(\beta)c_{\mu}(\alpha^{\prime}}}}\alpha) c_{\mu}(\underbracket{\beta)c_{\mu}(\alpha^{\prime})c_{\mu}(\beta^{\prime}})\rangle_{V}. 
\end{equation}
Each of the terms can be written as a product of two-point correlation functions. For example
\begin{equation}
\langle c_{\mu}(\overbracket{\alpha)c_{\mu}(\beta})c_{\mu}(\overbracket{\alpha^{\prime})c_{\mu}(\beta^{\prime}})\rangle_{V}=\langle c_{\mu}(\alpha)c_{\mu}(\beta)\rangle_{V}\langle c_{\mu}(\alpha^{\prime})c_{\mu}(\beta^{\prime})\rangle_{V}=\Lambda(\mu,\alpha)\delta_{\alpha\beta}\Lambda(\mu,\alpha^{\prime})\delta_{\alpha^{\prime}\beta^{\prime}}.
\end{equation}
The case for $\mu\neq\nu$ we need also to include the non-Gaussian corrections. We have 
\begin{equation}
\langle c_{\mu}(\alpha)c_{\nu}(\beta)c_{\mu}(\alpha^{\prime})c_{\nu}(\beta^{\prime})\rangle_{V}=
\langle c_{\mu}(\lefteqn{\overbracket{\phantom{\alpha)c_{\nu}(\beta)c_{\mu}(\alpha^{\prime}}}}\alpha) c_{\nu}(\underbracket{\beta)c_{\mu}(\alpha^{\prime})c_{\nu}(\beta^{\prime}})\rangle_{V}+
\langle c_{\mu}(\overbracket[2.0pt]{\alpha)c_{\nu}(\beta})c_{\mu}(\overbracket[2.0pt]{\alpha^{\prime})c_{\nu}(\beta^{\prime}})\rangle_{V}
+\langle c_{\mu}(\overbracket[2.0pt]{\alpha)c_{\nu}(\underbracket[2.0pt]{\beta)c_{\mu}(\alpha^{\prime}})c_{\nu}(\beta^{\prime}})\rangle_{V}.\label{NG1}
\end{equation}
The last two terms in (\ref{NG1}) arise as a result of the orthogonality condition between the many-body eigenstates. For example the first non-Gaussian term is 
\begin{equation}
\langle c_{\mu}(\overbracket[2.0pt]{\alpha)c_{\nu}(\beta})c_{\mu}(\overbracket[2.0pt]{\alpha^{\prime})c_{\nu}(\beta^{\prime}})\rangle_{V}
=-\frac{\Lambda(\mu,\alpha)\Lambda(\nu,\beta)\Lambda(\mu,\alpha^{\prime})\Lambda(\nu,\beta^{\prime})}{\sum_{\gamma}\Lambda(\mu,\gamma)\Lambda(\nu,\gamma)}\delta_{\alpha\beta}\delta_{\alpha^{\prime}\beta^{\prime}}
\end{equation}
and similarly for the second one.
\end{section}

\begin{section}{Calculation of the QFI using Random Matrix Approach}\label{QFI_C}
Here we provide the method which we use to evaluate the QFI (\ref{main}). We set $\partial_{\lambda}\hat{H}_{0}=\hat{H}_{0}^{\prime}$ and assume that $\hat{H}^{\prime}_{0}$ is a diagonal matrix in the non-interacting basis. Therefore, the QFI is 
\begin{eqnarray}
F_{Q}(\lambda)=4t^{2}\left\{\sum_{\mu\nu\rho}a^{*}_{\mu}a_{\nu}(\hat{H}^{\prime}_{0})_{\mu\rho}(\hat{H}^{\prime}_{0})_{\rho\nu}
e^{i\theta_{\mu\nu}t}{\rm sinc}(\theta_{\mu\rho}t)
{\rm sinc}(\theta_{\rho\nu}t)
-|\sum_{\mu\nu}a^{*}_{\mu}a_{\nu}e^{i\theta_{\mu\nu}t}(\hat{H}^{\prime}_{0})_{\mu\nu}{\rm sinc}(\theta_{\mu\nu}t)|^{2}\right\}.\label{fisher}
\end{eqnarray}
Let us now consider separately the first term in (\ref{fisher}), namely
\begin{eqnarray}
&&\sum_{\mu\nu\rho}a^{*}_{\mu}a_{\nu}\langle\psi_{\mu}|\hat{H}^{\prime}_{0}|\psi_{\rho}\rangle\langle\psi_{\rho}|\hat{H}^{\prime}_{0}|\psi_{\nu}\rangle e^{i\theta_{\mu\nu}t}
{\rm sinc}(\theta_{\mu\rho}t){\rm sinc}(\theta_{\rho\nu}t)=\sum_{\mu}|a_{\mu}|^{2}|\langle\psi_{\mu}|\hat{H}^{\prime}_{0}|\psi_{\mu}\rangle|^{2}+\sum_{\mu\nu\atop\mu\neq \nu}|a_{\mu}|^{2}|\langle\psi_{\mu}|\hat{H}^{\prime}_{0}|\psi_{\nu}\rangle|^{2}{\rm sinc}^{2}(\theta_{\mu\nu}t)\notag\\
&&+\sum_{\mu\nu\atop\mu\neq\nu}a^{*}_{\mu}a_{\nu}
\langle\psi_{\mu}|\hat{H}^{\prime}_{0}|\psi_{\nu}\rangle\langle\psi_{\nu}|\hat{H}^{\prime}_{0}|\psi_{\nu}\rangle e^{i\theta_{\mu\nu}t}{\rm sinc}(\theta_{\mu\nu}t)+\sum_{\mu\nu\atop\mu\neq\nu}a^{*}_{\mu}a_{\nu}\langle\psi_{\mu}|\hat{H}^{\prime}_{0}|\psi_{\mu}\rangle\langle\psi_{\mu}|\hat{H}^{\prime}_{0}|\psi_{\nu}\rangle e^{i\theta_{\mu\nu}t}{\rm sinc}(\theta_{\mu\nu}t)\notag\\
&&+\sum_{\mu\nu\rho\atop \mu\neq\nu\neq\rho}a^{*}_{\mu}a_{\nu}\langle\psi_{\mu}|\hat{H}^{\prime}_{0}|\psi_{\rho}\rangle\langle\psi_{\rho}|\hat{H}^{\prime}_{0}|\psi_{\nu}\rangle e^{i\theta_{\mu\nu}t}
{\rm sinc}(\theta_{\mu\rho}t){\rm sinc}(\theta_{\rho\nu}t).\label{firstT}
\end{eqnarray}
Now we apply the self-averaging condition for each of the terms in (\ref{firstT}). For the first one we have
\begin{equation}
\sum_{\mu}|a_{\mu}|^{2}|\langle\psi_{\mu}|\hat{H}^{\prime}_{0}|\psi_{\mu}\rangle|^{2}
=\sum_{\mu}\langle|a_{\mu}|^{2}|\langle\psi_{\mu}|\hat{H}^{\prime}_{0}|\psi_{\mu}\rangle|^{2}\rangle_{V}.\label{1}
\end{equation}
In order to evaluate (\ref{1}) we may further decouple the coefficients $a_{\mu}$ describing the initial state part and observable in the sense that (see Section "Corrections due to self-averaging decoupling" for more details)
\begin{equation}
\sum_{\mu}\langle|a_{\mu}|^{2}|\langle\psi_{\mu}|\hat{H}^{\prime}_{0}|\psi_{\mu}\rangle|^{2}\rangle_{V}
=\sum_{\mu}\langle |a_{\mu}|^{2}\rangle_{V}\langle|\langle \psi_{\mu}|\hat{H}^{\prime}_{0}|\psi_{\mu}\rangle|^{2}\rangle_{V}.
\end{equation}
Therefore, we have 
\begin{eqnarray}
\langle|\langle \psi_{\mu}|\hat{H}^{\prime}_{0}|\psi_{\mu}\rangle|^{2}\rangle_{V}&=&
\sum_{\alpha\beta}\langle c_{\mu}(\alpha)c_{\mu}(\alpha)c_{\mu}(\beta)c_{\mu}(\beta)\rangle_{V}(\hat{H}^{\prime}_{0})_{\alpha\alpha}(\hat{H}^{\prime}_{0})_{\beta\beta}\notag\\
&&=\sum_{\alpha\beta}\left(2\Lambda^{2}(\mu,\alpha)\delta_{\alpha\beta}(\hat{H}^{\prime}_{0})_{\alpha\alpha}
(\hat{H}^{\prime}_{0})_{\beta\beta}+\Lambda(\mu,\alpha)(\hat{H}^{\prime}_{0})_{\alpha\alpha}\Lambda(\mu,\beta)(\hat{H}^{\prime}_{0})_{\beta\beta}\right).
\end{eqnarray}

\begin{figure}
\includegraphics[width=0.48\textwidth]{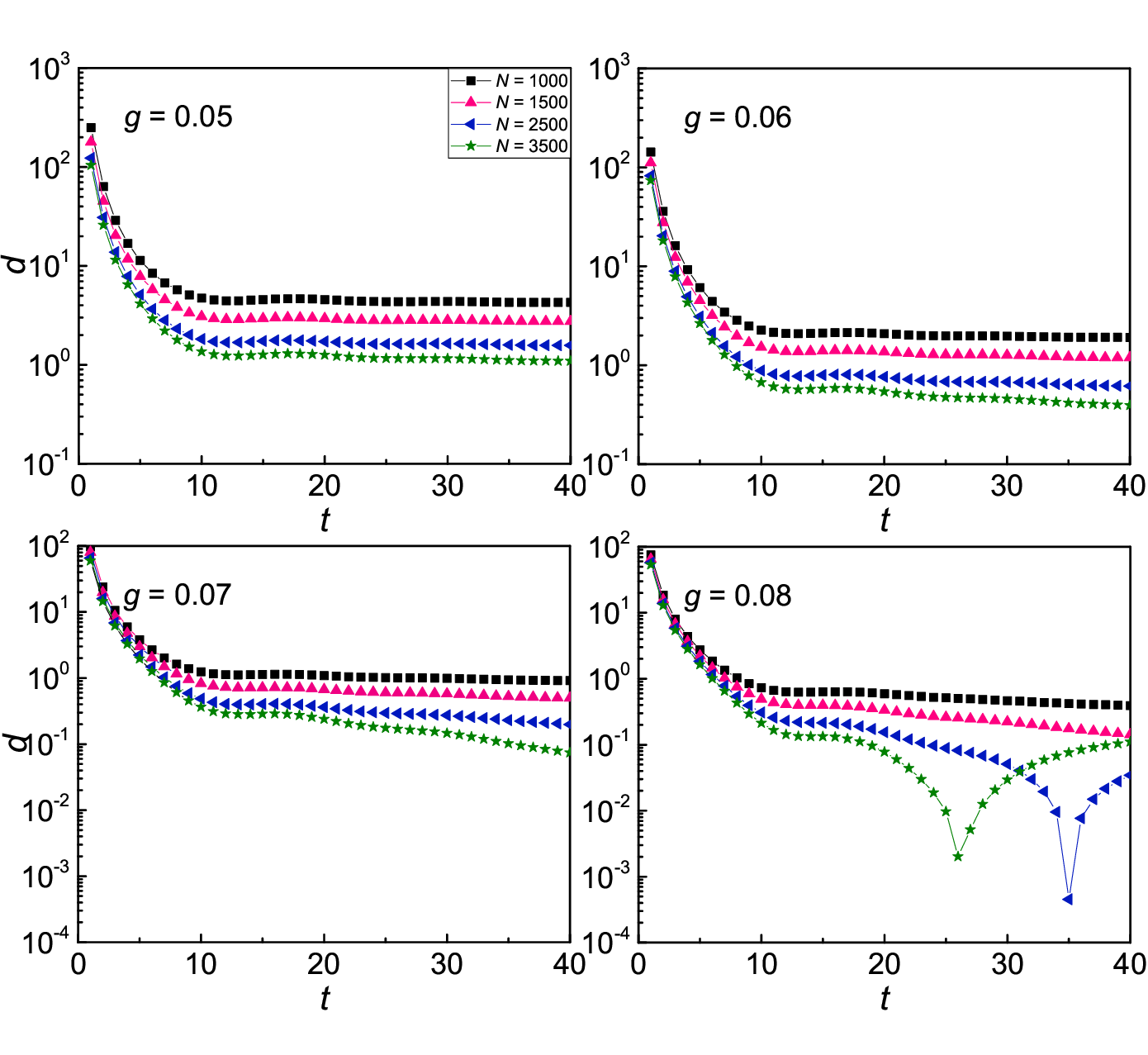}
\caption{Relative error $d=|1-(F_{Q}(\omega))_{\rm RMT}/(F_{Q}(\omega))|$ between results for the QFI derived from Eqs. (\ref{main}) and (\ref{QFI_def}).}
\label{fig_r_error}
\end{figure}
We define the average $[(\hat{H}_{0}^{\prime})_{\alpha\alpha}]_{\mu}=\sum_{\alpha}\Lambda(\mu,\alpha)(\hat{H}^{\prime}_{0})_{\alpha\alpha}$, which is essentially a microcanonical average centered on the energy $E_{\mu}$. We also apply the smoothness condition which implies that the variation of $[(\hat{H}_{0}^{\prime})_{\alpha\alpha}]_{\mu}$ as a function of $E_{\mu}$ can be neglected. Using this, we obtain
\begin{equation}
\langle|\langle \psi_{\mu}|\hat{H}^{\prime}_{0}|\psi_{\mu}\rangle|^{2}\rangle_{V}\approx
2[(\hat{H}^{\prime}_{0})^{2}_{\alpha\alpha}]_{\mu}\sum_{\alpha}\Lambda^{2}(\mu,\alpha)+[(\hat{H}^{\prime}_{0})_{\alpha\alpha}]^{2}_{\mu}.
\end{equation}
Further, we take the continuum limit, substituting $\sum_{\alpha}\rightarrow \int_{-\infty}^{\infty}\frac{dE_{\alpha}}{\omega}$, and thereby obtain
\begin{equation}
\sum_{\alpha}\Lambda^{2}(\mu,\alpha)=\frac{1}{\omega}\int_{-\infty}^{\infty}\Lambda^{2}(\mu,\alpha)d E_{\alpha}=\frac{\omega}{2\pi\Gamma}.
\end{equation}
Substituting in Eq. (\ref{1}) we have
\begin{equation}
\sum_{\mu}|a_{\mu}|^{2}|\langle\psi_{\mu}|\hat{H}^{\prime}_{0}|\psi_{\mu}\rangle|^{2}
=\sum_{\mu}\langle|a_{\mu}|^{2}\rangle_{V}\left(\frac{\omega}{\pi\Gamma}[(\hat{H}^{\prime}_{0})^{2}_{\alpha\alpha}]_{\mu}
+[(\hat{H}^{\prime}_{0})_{\alpha\alpha}]^{2}_{\mu}\right).\label{1new}
\end{equation}
As long as $[(\hat{H}^{\prime}_{0})^{2}_{\alpha\alpha}]_{\mu}$ and $[(\hat{H}^{\prime}_{0})_{\alpha\alpha}]^{2}_{\mu}$ are smooth function of the energy $E_{\mu}$ and the probabilities $|a_{\mu}|^{2}$ take non-vanishing value close to the mean energy $E_{0}=\langle\Psi_{0}|\hat{H}|\Psi_{0}\rangle$ with $|\Psi_{0}\rangle$ being the initial state, the ETH ensures that Eq. (\ref{1new}) is equivalent to a microcanonical average,
\begin{equation}
\sum_{\mu}|a_{\mu}|^{2}|\langle\psi_{\mu}|\hat{H}^{\prime}_{0}|\psi_{\mu}\rangle|^{2}
=\frac{\omega}{\pi\Gamma}(\hat{H}^{\prime 2}_{0})_{\rm mc}+(\hat{H}^{\prime}_{0})_{\rm mc}^{2}.\label{a1}
\end{equation}
Consider the second term in (\ref{firstT})
\begin{equation}
\sum_{\mu\nu\atop\mu\neq \nu}|a_{\mu}|^{2}|\langle\psi_{\mu}|\hat{H}^{\prime}_{0}|\psi_{\nu}\rangle|^{2}{\rm sinc}^{2}(\theta_{\mu\nu}t)=
\sum_{\mu\nu\atop\mu\neq \nu}\langle|a_{\mu}|^{2}\rangle_{V}\langle|\langle\psi_{\mu}|\hat{H}^{\prime}_{0}|\psi_{\nu}\rangle|^{2}\rangle_{V}{\rm sinc}^{2}(\theta_{\mu\nu}t).
\end{equation}
For the matrix element we have
\begin{equation}
\langle|\langle\psi_{\mu}|\hat{H}^{\prime}_{0}|\psi_{\nu}\rangle|^{2}\rangle_{V}
=\sum_{\alpha\beta}\langle c_{\mu}(\alpha)c_{\nu}(\alpha)c_{\mu}(\beta)c_{\nu}(\beta)\rangle_{V}(\hat{H}^{\prime}_{0})_{\alpha\alpha}(\hat{H}^{\prime}_{0})_{\beta\beta}.
\end{equation}
Now, using (\ref{NG}) we obtain
\begin{equation}
\langle c_{\mu}(\alpha)c_{\nu}(\alpha)c_{\mu}(\beta)c_{\nu}(\beta)\rangle_{V}=
\Lambda(\mu,\alpha)\Lambda(\nu,\alpha)\delta_{\alpha\beta}-\frac{\Lambda(\mu,\alpha)\Lambda(\nu,\alpha)\Lambda(\mu,\beta)\Lambda(\nu,\beta)}{\sum_{\gamma}\Lambda(\mu,\gamma)\Lambda(\nu,\gamma)}-
\frac{\Lambda^{2}(\mu,\alpha)\Lambda^{2}(\nu,\alpha)}{\sum_{\gamma}\Lambda(\mu,\gamma)\Lambda(\nu,\gamma)}\delta_{\alpha\beta}
\end{equation}
and the matrix element becomes
\begin{eqnarray}\label{eq:off_diag_integral}
\langle|\langle\psi_{\mu}|\hat{H}^{\prime}_{0}|\psi_{\nu}\rangle|^{2}\rangle_{V}&=&\sum_{\alpha\beta}
\left(\Lambda(\mu,\alpha)\Lambda(\nu,\alpha)\delta_{\alpha\beta}-\frac{\Lambda(\mu,\alpha)\Lambda(\nu,\alpha)\Lambda(\mu,\beta)\Lambda(\nu,\beta)}{\sum_{\gamma}\Lambda(\mu,\gamma)\Lambda(\nu,\gamma)}-
\frac{\Lambda^{2}(\mu,\alpha)\Lambda^{2}(\nu,\alpha)}{\sum_{\gamma}\Lambda(\mu,\gamma)\Lambda(\nu,\gamma)}\delta_{\alpha\beta}\right)(\hat{H}^{\prime}_{0})_{\alpha\alpha}(\hat{H}^{\prime}_{0})_{\beta\beta}\notag\\
&&\approx [(\Delta\hat{H}^{\prime}_{0})_{\alpha\alpha}]^{2}_{\bar{\mu}}\sum_{\alpha}\Lambda(\mu,\alpha)\Lambda(\nu,\alpha),\label{4P}
\end{eqnarray}
where $[(\Delta\hat{H}^{\prime}_{0})_{\alpha\alpha}]^{2}_{\bar{\mu}}$ is the variance, $\bar{\mu}=\frac{\mu+\nu}{2}$. Going in the continuum limit we get
\begin{equation}
\sum_{\alpha}\Lambda(\mu,\alpha)\Lambda(\nu,\alpha)=\frac{2\omega\Gamma}{\pi}\frac{1}{(E_{\mu}-E_{\nu})^{2}+4\Gamma^{2}}.\label{int}
\end{equation}
Therefore, the second term becomes
\begin{eqnarray}
&&\sum_{\mu\nu\atop\mu\neq \nu}\langle|a_{\mu}|^{2}\rangle_{V}\langle|\langle\psi_{\mu}|\hat{H}^{\prime}_{0}|\psi_{\nu}\rangle|^{2}\rangle_{V}{\rm sinc}^{2}(\theta_{\mu\nu}t)=\frac{2\omega\Gamma}{\pi}\sum_{\mu\nu\atop\mu\neq \nu}\langle|a_{\mu}|^{2}\rangle_{V}[(\Delta\hat{H}^{\prime}_{0})_{\alpha\alpha}]^{2}_{\bar{\mu}}\frac{{\rm sinc}^{2}(\theta_{\mu\nu}t)}{(E_{\mu}-E_{\nu})^{2}+4\Gamma^{2}}.\label{linear}
\end{eqnarray}
We now replace the sum over the index $\nu$ with integration, namely
\begin{equation}
\sum_{\nu}\frac{{\rm sinc}^{2}(\theta_{\mu\nu}t)}{(E_{\mu}-E_{\nu})^{2}+4\Gamma^{2}}\rightarrow \frac{1}{\omega}
\int_{-\infty}^{\infty}\frac{{\rm sinc}^{2}(\theta_{\mu\nu}t)}{(E_{\mu}-E_{\nu})^{2}+4\Gamma^{2}}dE_{\nu}=\frac{t}{2\omega}
\int_{-\infty}^{\infty}\frac{{\rm sinc}^{2}(x)}{x^{2}+(\Gamma t)^{2}}dx=\frac{\pi t}{4\omega(\Gamma t)^{3}}(e^{-2\Gamma t}-1+2\Gamma t).
\end{equation}
Finally, we have
\begin{equation}
\sum_{\mu\nu\atop\mu\neq \nu}|a_{\mu}|^{2}|\langle\psi_{\mu}|\hat{H}^{\prime}_{0}|\psi_{\nu}\rangle|^{2}{\rm sinc}^{2}(\theta_{\mu\nu}t)=\frac{(\Delta\hat{H}^{\prime 2}_{0})_{\rm mc}}{2(\Gamma t)^{2}}(e^{-2\Gamma t}-1+2\Gamma t).
\end{equation}
We note that in Eq. (\ref{4P}) we have neglected the contribution from the third term which is of order of $(\omega/\Gamma)^{2}$. Indeed, we have 
\begin{eqnarray}
&&\sum_{\alpha}\Lambda^{2}(\mu,\alpha)\Lambda^{2}(\nu,\alpha)\rightarrow
\left(\frac{\omega\Gamma}{\pi}\right)^{4}\frac{1}{\omega}\int_{-\infty}^{\infty}
\frac{dE_{\alpha}}{((E_{\mu}-E_{\alpha})^{2}+\Gamma^{2})^{2}((E_{\nu}-E_{\alpha})^{2}+\Gamma^{2})^{2}}\notag\\
&&=\left(\frac{\omega\Gamma}{\pi}\right)^{4}\frac{\pi}{\omega\Gamma^{3}}\frac{(E_{\mu}-E_{\nu})^{2}+20\Gamma^{2}}{((E_{\mu}-E_{\nu})^{2}+4\Gamma^{2})^{3}}.
\end{eqnarray}
Therefore, using (\ref{int}) we conclude that the third term in (\ref{4P}) is of order of $(\omega/\Gamma)^{2}$. We also note that the contribution for $\mu=\nu$ in Eq. (\ref{linear}) can be neglected.

Consider the third term in (\ref{firstT}). We have
\begin{equation}
\sum_{\mu\nu\atop\mu\neq\nu}a^{*}_{\mu}a_{\nu}
\langle\psi_{\mu}|\hat{H}^{\prime}_{0}|\psi_{\nu}\rangle\langle\psi_{\nu}|\hat{H}^{\prime}_{0}|\psi_{\nu}\rangle e^{i\theta_{\mu\nu}t}{\rm sinc}(\theta_{\mu\nu}t)=\sum_{\mu\nu\atop\mu\neq\nu}\langle a^{*}_{\mu}a_{\nu}\rangle_{V}\langle \langle\psi_{\mu}|\hat{H}^{\prime}_{0}|\psi_{\nu}\rangle\langle\psi_{\nu}|\hat{H}^{\prime}_{0}|\psi_{\nu}\rangle \rangle_{V}e^{i\theta_{\mu\nu}t}{\rm sinc}(\theta_{\mu\nu}t).
\end{equation}
The matrix elements express in the non-interacting basis are
\begin{equation}
\langle \langle\psi_{\mu}|\hat{H}^{\prime}_{0}|\psi_{\nu}\rangle\langle\psi_{\nu}|\hat{H}^{\prime}_{0}|\psi_{\nu}\rangle \rangle_{V}=
\sum_{\alpha\beta}\langle c_{\mu}(\alpha)c_{\nu}(\alpha)c_{\nu}(\beta)c_{\nu}(\beta)\rangle_{V}(\hat{H}^{\prime}_{0})_{\alpha\alpha}(\hat{H}^{\prime}_{0})_{\beta\beta}.
\end{equation}
Using the generating function Eq. (\ref{GF2}) the four-point correlation function is
\begin{equation}
\langle c_{\mu}(\alpha)c_{\nu}(\alpha)c_{\nu}(\beta)c_{\nu}(\beta)\rangle_{V}=0.
\end{equation}
Similarly, for the last term in  (\ref{firstT}) we have
\begin{equation}
\langle\langle\psi_{\mu}|\hat{H}^{\prime}_{0}|\psi_{\rho}\rangle\langle\psi_{\rho}|\hat{H}^{\prime}_{0}|\psi_{\nu}\rangle\rangle_{V}=
\sum_{\alpha\beta}\langle c_{\mu}(\alpha)c_{\rho}(\alpha)c_{\rho}(\beta)c_{\nu}(\beta)\rangle_{V}(\hat{H}^{\prime}_{0})_{\alpha\alpha}(\hat{H}^{\prime}_{0})_{\beta\beta}.
\end{equation}
Now, the four-point correlation function contains three different indices. In order to evaluate it we need to introduce three auxiliary fields. Because the indexes $\mu$ and $\nu$ repeat only once we have 
\begin{equation}
\langle c_{\mu}(\alpha)c_{\rho}(\alpha)c_{\rho}(\beta)c_{\nu}(\beta)\rangle_{V}=0.
\end{equation}
Combining all averages in (\ref{firstT}) we obtain
\begin{eqnarray}
\sum_{\mu\nu\rho}a^{*}_{\mu}a_{\nu}\langle\psi_{\mu}|\hat{H}^{\prime}_{0}|\psi_{\rho}\rangle\langle\psi_{\rho}|\hat{H}^{\prime}_{0}|\psi_{\nu}\rangle e^{i\theta_{\mu\nu}t}
{\rm sinc}(\theta_{\mu\rho}t){\rm sinc}(\theta_{\rho\nu}t)&=&
\frac{\omega}{\pi\Gamma}(\hat{H}^{\prime 2}_{0})_{\rm mc}+(\hat{H}^{\prime}_{0})_{\rm mc}^{2}+\frac{(\Delta\hat{H}^{\prime 2}_{0})_{\rm mc}}{2(\Gamma t)^{2}}(e^{-2\Gamma t}-1+2\Gamma t).\label{final_1}
\end{eqnarray}

Let us now consider separately the second term in (\ref{fisher}), namely
\begin{eqnarray}
&&\sum_{\mu\nu}\sum_{\mu^{\prime}\nu^{\prime}}a^{*}_{\mu}a_{\nu}a_{\mu^{\prime}}a^{*}_{\nu^{\prime}}\langle\psi_{\mu}|\hat{H}^{\prime}_{0}|\psi_{\nu}\rangle
\langle\psi_{\nu^{\prime}}|\hat{H}^{\prime}_{0}|\psi_{\mu^{\prime}}\rangle e^{i\theta_{\mu\nu}t}e^{-i\theta_{\mu^{\prime}\nu^{\prime}}t}{\rm sinc}(\theta_{\mu\nu}t){\rm sinc}(\theta_{\mu^{\prime}\nu^{\prime}}t)=\sum_{\mu}|a_{\mu}|^{4}|\langle\psi_{\mu}|\hat{H}^{\prime}_{0}|\psi_{\mu}\rangle|^{2}\notag\\
&&+\sum_{\mu\nu\atop\mu\neq\nu}|a_{\mu}|^{2}|a_{\nu}|^{2}
\langle\psi_{\mu}|\hat{H}^{\prime}_{0}|\psi_{\mu}\rangle\langle\psi_{\nu}|\hat{H}^{\prime}_{0}|\psi_{\nu}\rangle+\sum_{\mu\nu\atop\mu\neq\nu}|a_{\mu}|^{2}|a_{\nu}|^{2}
\langle\psi_{\mu}|\hat{H}^{\prime}_{0}|\psi_{\nu}\rangle\langle\psi_{\nu}|\hat{H}^{\prime}_{0}|\psi_{\mu}\rangle {\rm sinc}^{2}(\theta_{\mu\nu}t)\notag\\
&&+\sum_{\mu\nu\atop\mu\neq\nu}a_{\mu}^{*2}a_{\nu}^{2}
\langle\psi_{\mu}|\hat{H}^{\prime}_{0}|\psi_{\nu}\rangle\langle\psi_{\mu}|\hat{H}^{\prime}_{0}|\psi_{\nu}\rangle e^{2i\theta_{\mu\nu}t}{\rm sinc}^{2}(\theta_{\mu\nu}t)+\ldots\label{second}
\end{eqnarray}
We have
\begin{equation}
\sum_{\mu}|a_{\mu}|^{4}|\langle\psi_{\mu}|\hat{H}^{\prime}_{0}|\psi_{\mu}\rangle|^{2}=\sum_{\mu}\langle|a_{\mu}|^{4}\rangle_{V}\left(\frac{\omega}{\pi\Gamma}[(\hat{H}^{\prime}_{0})^{2}_{\alpha\alpha}]_{\mu}
+[(\hat{H}^{\prime}_{0})_{\alpha\alpha}]^{2}_{\mu}\right)=
\left(\frac{\omega}{\pi\Gamma}(\hat{H}^{\prime 2}_{0})_{\rm mc}
+(\hat{H}^{\prime}_{0})^{2}_{\rm mc}\right)\sum_{\mu}\langle|a_{\mu}|^{4}\rangle_{V}.
\end{equation}
Note that $\left(\frac{\omega}{\pi\Gamma}\right)(\hat{H}^{\prime 2}_{0})_{\rm mc}\langle\sum_{\mu}|a_{\mu}|^{4}\rangle_{V}$ is of order of $(\omega/\Gamma)^{2}$ and thereby we neglect. The second term in (\ref{second}) is 
\begin{equation}
\sum_{\mu\nu\atop\mu\neq\nu}|a_{\mu}|^{2}|a_{\nu}|^{2}
\langle\psi_{\mu}|\hat{H}^{\prime}_{0}|\psi_{\mu}\rangle\langle\psi_{\nu}|\hat{H}^{\prime}_{0}|\psi_{\nu}\rangle=
\sum_{\mu\nu\atop\mu\neq\nu}\langle|a_{\mu}|^{2}|a_{\nu}|^{2}\rangle_{V}\langle 
\langle\psi_{\mu}|\hat{H}^{\prime}_{0}|\psi_{\mu}\rangle\langle\psi_{\nu}|\hat{H}^{\prime}_{0}|\psi_{\nu}\rangle\rangle_{V}.
\end{equation}
The matrix elements are 
\begin{eqnarray}
\langle\langle\psi_{\mu}|\hat{H}^{\prime}_{0}|\psi_{\mu}\rangle\langle\psi_{\nu}|\hat{H}^{\prime}_{0}|\psi_{\nu}\rangle\rangle_{V}&=&
\sum_{\alpha\beta}\langle c_{\mu}(\alpha)c_{\mu}(\alpha)c_{\nu}(\beta)c_{\nu}(\beta)\rangle_{V}(\hat{H}_{0}^{\prime})_{\alpha\alpha}
(\hat{H}_{0}^{\prime})_{\beta\beta}\notag\\
&&=\sum_{\alpha\beta}\left(\Lambda(\mu,\alpha)\Lambda(\nu,\beta)-2\frac{\Lambda^{2}(\mu,\alpha)\Lambda^{2}(\nu,\beta)}{\sum_{\gamma}\Lambda(\mu,\gamma)\Lambda(\nu,\gamma)}\delta_{\alpha\beta}\right)
(\hat{H}_{0}^{\prime})_{\alpha\alpha}
(\hat{H}_{0}^{\prime})_{\beta\beta}\notag\\
&&\approx [(\hat{H}^{\prime}_{0})_{\alpha\alpha}]^{2}_{\bar{\mu}},
\end{eqnarray}
where we neglect the second term which is of order of $(\omega/\Gamma)^{2}$. Therefore, we obtain
\begin{equation}
\sum_{\mu\nu\atop\mu\neq\nu}\langle|a_{\mu}|^{2}|a_{\nu}|^{2}\rangle_{V}[(\hat{H}^{\prime}_{0})_{\alpha\alpha}]^{2}_{\bar{\mu}}\approx (\hat{H}^{\prime}_{0})_{\rm mc}^{2}
(\sum_{\mu\nu}\langle |a_{\mu}|^{2}|a_{\nu}|^{2}\rangle_{V}-\sum_{\mu}\langle |a_{\mu}|^{4}\rangle_{V})
=(\hat{H}^{\prime}_{0})_{\rm mc}^{2}(1-\sum_{\mu}\langle |a_{\mu}|^{4}\rangle_{V}).
\end{equation}

Consider the term
\begin{equation}
\sum_{\mu\nu\atop\mu\neq\nu}|a_{\mu}|^{2}|a_{\nu}|^{2}
\langle|\langle\psi_{\mu}|\hat{H}^{\prime}_{0}|\psi_{\nu}\rangle|^{2}\rangle_{V} {\rm sinc}^{2}(\theta_{\mu\nu}t)=
\sum_{\mu\nu\atop\mu\neq\nu}\langle|a_{\mu}|^{2}|a_{\nu}|^{2}\rangle_{V}
\langle|\langle\psi_{\mu}|\hat{H}^{\prime}_{0}|\psi_{\nu}\rangle|^{2}\rangle_{V} {\rm sinc}^{2}(\theta_{\mu\nu}t).
\end{equation}
Using (\ref{4P}) we obtain
\begin{equation}
\sum_{\mu\nu\atop\mu\neq\nu}\langle|a_{\mu}|^{2}|a_{\nu}|^{2}\rangle_{V}
\langle|\langle\psi_{\mu}|\hat{H}^{\prime}_{0}|\psi_{\nu}\rangle|^{2}\rangle_{V} {\rm sinc}^{2}(\theta_{\mu\nu}t)\approx 
(\Delta\hat{H}^{\prime 2}_{0})_{\rm mc}\frac{2\Gamma\omega}{\pi}
\sum_{\mu\nu\atop\mu\neq\nu}\langle|a_{\mu}|^{2}|a_{\nu}|^{2}\rangle_{V}\frac{{\rm sinc}^{2}(\theta_{\mu\nu}t)}{(E_{\mu}-E_{\nu})^{2}+4\Gamma^{2}}.
\end{equation}
Let us now assume that the initial state is an eigenstate of the non-interaction Hamiltonian $\hat{H}_{0}$, namely
$|\psi(0)\rangle=|\varphi_{\alpha_{0}}\rangle$. Then, we have
\begin{equation}
\langle|a_{\mu}|^{2}|a_{\nu}|^{2}\rangle_{V}=\Lambda(\mu,\alpha_{0})\Lambda(\nu,\alpha_{0})
\end{equation}
and we get
\begin{equation}
(\Delta\hat{H}^{\prime 2}_{0})_{\rm mc}\frac{2\Gamma\omega}{\pi}
\sum_{\mu\nu\atop\mu\neq\nu}\langle|a_{\mu}|^{2}|a_{\nu}|^{2}\rangle_{V}\frac{{\rm sinc}^{2}(\theta_{\mu\nu}t)}{(E_{\mu}-E_{\nu})^{2}+4\Gamma^{2}}=(\Delta\hat{H}^{\prime 2}_{0})_{\rm mc}\frac{2\Gamma\omega}{\pi}
\sum_{\mu\nu\atop\mu\neq\nu}\Lambda(\mu,\alpha_{0})\Lambda(\nu,\alpha_{0})\frac{{\rm sinc}^{2}(\theta_{\mu\nu}t)}{(E_{\mu}-E_{\nu})^{2}+4\Gamma^{2}}.
\end{equation}
We replace the sum with the integration, such that we have
\begin{equation}
\frac{2\Gamma\omega}{\pi}\frac{1}{\omega}\int_{-\infty}^{\infty}\Lambda(\mu,\alpha_{0})\frac{{\rm sinc}^{2}(\theta_{\mu\nu}t)}{(E_{\mu}-E_{\nu})^{2}+4\Gamma^{2}}dE_{\mu}\leq \frac{2\Gamma}{\pi}\frac{\omega}{\pi\Gamma}
\int_{-\infty}^{\infty}\frac{{\rm sinc}^{2}(\theta_{\mu\nu}t)}{(E_{\mu}-E_{\nu})^{2}+4\Gamma^{2}}dE_{\mu}
=\frac{\omega}{\pi\Gamma}\frac{1}{2(\Gamma t)^{2}}(e^{-2\Gamma t}-1+2\Gamma t).
\end{equation}
In the above equation we have used that for any two functions $f(x)>0$ and $g(x)>0$, which obey $f(x)g(x)\leq f_{\rm max}g(x)$ it follow that $\int_{-\infty}^{\infty}f(x)g(x)dx\leq f_{\rm max}\int_{-\infty}^{\infty}g(x)dx$. Therefore we obtain
\begin{equation}
\sum_{\mu\nu\atop\mu\neq\nu}\langle|a_{\mu}|^{2}|a_{\nu}|^{2}\rangle_{V}
\langle|\langle\psi_{\mu}|\hat{H}^{\prime}_{0}|\psi_{\nu}\rangle|^{2}\rangle_{V} {\rm sinc}^{2}(\theta_{\mu\nu}t)\leq
(\Delta\hat{H}^{\prime 2}_{0})_{\rm mc}\frac{\omega}{\pi\Gamma}\frac{1}{2(\Gamma t)^{2}}(e^{-2\Gamma t}-1+2\Gamma t).
\end{equation}
As long as $\Gamma\gg\omega$ we neglect this term.

Similarly, we have 
\begin{equation}
2\sum_{\mu\nu\atop\mu\neq\nu}a_{\mu}^{*2}a_{\nu}^{2}
|\langle\psi_{\mu}|\hat{H}^{\prime}_{0}|\psi_{\nu}\rangle|^{2}\cos(2\theta_{\mu\nu}t){\rm sinc}^{2}(\theta_{\mu\nu}t)\approx 2(\Delta\hat{H}^{\prime 2}_{0})_{\rm mc}\frac{2\Gamma\omega}{\pi}
\sum_{\mu\nu\atop\mu\neq\nu}\Lambda(\mu,\alpha_{0})\Lambda(\nu,\alpha_{0})\frac{\cos(2\theta_{\mu\nu}t)
{\rm sinc}^{2}(\theta_{\mu\nu}t)}{(E_{\mu}-E_{\nu})^{2}+4\Gamma^{2}}.
\end{equation}
Replacing the sum with integration we get
\begin{equation}
\frac{2\Gamma\omega}{\pi}\int_{-\infty}^{\infty}\Lambda(\mu,\alpha_{0})\frac{\cos(2\theta_{\mu\nu}t)
{\rm sinc}^{2}(\theta_{\mu\nu}t)}{(E_{\mu}-E_{\nu})^{2}+4\Gamma^{2}}dE_{\mu}\leq \frac{2\Gamma}{\pi}\frac{\omega}{\pi\Gamma}
\int_{-\infty}^{\infty}\frac{\cos(2\theta_{\mu\nu}t)
{\rm sinc}^{2}(\theta_{\mu\nu}t)}{(E_{\mu}-E_{\nu})^{2}+4\Gamma^{2}}dE_{\mu}.
\end{equation}
The integral is given by
\begin{equation}
\frac{2\Gamma}{\pi}\int_{-\infty}^{\infty}\frac{\cos(2\theta_{\mu\nu}t)
{\rm sinc}^{2}(\theta_{\mu\nu}t)}{(E_{\mu}-E_{\nu})^{2}+4\Gamma^{2}}dE_{\mu}=\frac{(e^{-2\Gamma t}-1)^{2}}{4(\Gamma t)^{2}}.
\end{equation}
Therefore we obtain
\begin{equation}
2\sum_{\mu\nu\atop\mu\neq\nu}a_{\mu}^{*2}a_{\nu}^{2}
\langle\psi_{\mu}|\hat{H}^{\prime}_{0}|\psi_{\nu}\rangle\langle\psi_{\mu}|\hat{H}^{\prime}_{0}|\psi_{\nu}\rangle \cos(2\theta_{\mu\nu}t){\rm sinc}^{2}(\theta_{\mu\nu}t)\leq 
(\Delta\hat{H}^{\prime 2}_{0})_{\rm mc}\left(\frac{\omega}{\pi\Gamma}\right)\frac{(e^{-2\Gamma t}-1)^{2}}{4(\Gamma t)^{2}},
\end{equation}
which we neglect in the limit $\Gamma\gg\omega$. All other terms in (\ref{second}) contain matrix elements with two and three equal indexes and respectively four different indexes and their ensemble average is zero.

Combining all averages in the second term (\ref{second}) we get

\begin{equation}
\sum_{\mu\nu}\sum_{\mu^{\prime}\nu^{\prime}}a^{*}_{\mu}a_{\nu}a_{\mu^{\prime}}a^{*}_{\nu^{\prime}}\langle\psi_{\mu}|\hat{H}^{\prime}_{0}|\psi_{\nu}\rangle
\langle\psi_{\nu^{\prime}}|\hat{H}^{\prime}_{0}|\psi_{\mu^{\prime}}\rangle e^{i\theta_{\mu\nu}t}e^{-i\theta_{\mu^{\prime}\nu^{\prime}}t}{\rm sinc}(\theta_{\mu\nu}t){\rm sinc}(\theta_{\mu^{\prime}\nu^{\prime}}t)\approx
(\hat{H}^{\prime}_{0})^{2}_{\rm mc}.\label{final_2}
\end{equation}
Finally, using Eq. (\ref{fisher}), (\ref{final_1}), and (\ref{final_2}) we obtain
\begin{equation}
F_{Q}(\lambda)=4t^{2}\left\{\frac{\omega}{\pi\Gamma}(\hat{H}^{\prime2}_{0})_{\rm mc}+\frac{(\Delta \hat{H}_{0}^{\prime2})_{\rm mc}}{2(\Gamma t)^{2}}
(e^{-2\Gamma t}-1+2\Gamma t)\right\}.\label{main1}
\end{equation}
In Fig. (\ref{fig_r_error}) we show the relative error between the exact result derived from Eq. (\ref{Fisher}) and the analytical formula Eq. (\ref{main1}) for various $g$ and $N$.

We see that increasing time the scaling of the QFI passes from linear to quadratic. In fact we can obtain the long time scaling of QFI using that $\lim_{t\rightarrow\infty}{\rm sinc}(x)=0$. Using (\ref{fisher}) we obtain
\begin{eqnarray}
\lim_{t\rightarrow\infty}\frac{F_{Q}(\lambda)}{t^{2}}&=&4\{\sum_{\mu}|a_{\mu}|^{2}|\langle\psi_{\mu}|\hat{H}^{\prime}_{0}|\psi_{\mu}\rangle|^{2}-\sum_{\mu\nu}|a_{\mu}|^{2}|a_{\nu}|^{2}\langle\psi_{\mu}|\hat{H}^{\prime}_{0}|\psi_{\mu}\rangle\langle\psi_{\nu}|\hat{H}^{\prime}_{0}|\psi_{\nu}\rangle\}\notag\\
&&=4\{\sum_{\mu}|a_{\mu}|^{2}|\langle\psi_{\mu}|\hat{H}^{\prime}_{0}|\psi_{\mu}\rangle|^{2}-\sum_{\mu}|a_{\mu}|^{4}|\langle\psi_{\mu}|\hat{H}^{\prime}_{0}|\psi_{\mu}\rangle|^{2}-
\sum_{\mu\nu\atop\mu\neq\nu}|a_{\mu}|^{2}|a_{\nu}|^{2}\langle\psi_{\mu}|\hat{H}^{\prime}_{0}|\psi_{\mu}\rangle\langle\psi_{\nu}|\hat{H}^{\prime}_{0}|\psi_{\nu}\rangle\}\notag\\
&&=4\{\frac{\omega}{\pi\Gamma}(\hat{H}^{\prime 2}_{0})_{\rm mc}+(\hat{H}^{\prime}_{0})_{\rm mc}^{2}
-(\hat{H}^{\prime}_{0})^{2}_{\rm mc}\sum_{\mu}\langle|a_{\mu}|^{4}\rangle_{V}-(\hat{H}^{\prime}_{0})_{\rm mc}^{2}(1-\sum_{\mu}\langle |a_{\mu}|^{4}\rangle_{V})\}.
\end{eqnarray}
Therefore, neglecting terms of order of $(\omega/\Gamma)^{2}$ we obtain the long time limit of QFI as $F_{Q}(\lambda)\approx (\hat{H}^{\prime 2}_{0})_{\rm mc}(4\omega/\pi\Gamma)t^{2}$. Similarly, we may consider the short-time limit of the QFI by using that $\lim_{t\rightarrow 0}{\rm sinc}(x)=1$. Then we have
\begin{eqnarray}
\lim_{t\rightarrow 0}\frac{F_{Q}(\lambda)}{t^{2}}&=&4\{\sum_{\mu}|a_{\mu}|^{2}|\langle\psi_{\mu}|\hat{H}^{\prime}_{0}|\psi_{\mu}\rangle|^{2}+\sum_{\mu\nu\atop\mu\neq \nu}|a_{\mu}|^{2}|\langle\psi_{\mu}|\hat{H}^{\prime}_{0}|\psi_{\nu}\rangle|^{2}
-\sum_{\mu}|a_{\mu}|^{4}|\langle\psi_{\mu}|\hat{H}^{\prime}_{0}|\psi_{\mu}\rangle|^{2}\notag\\
&&-
\sum_{\mu\nu\atop\mu\neq\nu}|a_{\mu}|^{2}|a_{\nu}|^{2}\langle\psi_{\mu}|\hat{H}^{\prime}_{0}|\psi_{\mu}\rangle\langle\psi_{\nu}|\hat{H}^{\prime}_{0}|\psi_{\nu}\rangle\}=4\{\frac{\omega}{\pi\Gamma}
(\hat{H}^{\prime 2}_{0})_{\rm mc}+\frac{2\omega\Gamma}{\pi}(\Delta \hat{H}_{0}^{\prime2})_{\rm mc}
\sum_{\mu\nu\atop\mu\neq\nu}\frac{|a_{\mu}|^{2}}{(E_{\mu}-E_{\nu})^{2}+4\Gamma^{2}}\}.
\end{eqnarray}
Furthermore, we replace the sum with integration such that we have,
\begin{equation}
\sum_{\nu}\frac{1}{(E_{\mu}-E_{\nu})^{2}+4\Gamma^{2}}\rightarrow\frac{1}{\omega}
\int_{-\infty}^{\infty}\frac{dE_{\nu}}{(E_{\mu}-E_{\nu})^{2}+4\Gamma^{2}}=\frac{\pi}{2\omega\Gamma}.
\end{equation}
Therefore, neglecting the terms of order of $\omega/\Gamma$, the short-time scaling of the QFI is $F_{Q}(\lambda)\approx4t^{2}(\Delta \hat{H}_{0}^{\prime2})_{\rm mc}$.
\end{section}

\begin{section}{Corrections due to self-averaging decoupling}\label{decoupling}

Let us assume that the initial state is an eigenstate of non-interaction Hamiltonian $\hat{H}_{0}$, namely $|\Psi_{0}\rangle=|\varphi_{\alpha_{0}}\rangle$. Therefore, we have 
\begin{equation}
\sum_{\mu}\langle|a_{\mu}|^{2}|\langle\psi_{\mu}|\hat{H}^{\prime}_{0}|\psi_{\mu}\rangle|^{2}\rangle_{V}
=\sum_{\mu}\sum_{\alpha\beta}\langle c_{\mu}(\alpha_{0})c_{\mu}(\alpha_{0})c_{\mu}(\alpha)c_{\mu}(\alpha)c_{\mu}(\beta)c_{\mu}(\beta)\rangle_{V}(\hat{H}^{\prime}_{0})_{\alpha\alpha}(\hat{H}^{\prime}_{0})_{\beta\beta}.
\end{equation}
In order to evaluate the average we use the generating function (\ref{CF11}). There are in total $15$ terms. Consider the term
\begin{equation}
\langle c_{\mu}(\overbracket{\alpha_{0})c_{\mu}(\alpha_{0}}) c_{\mu}(\lefteqn{\overbracket{\phantom{\alpha)c_{\mu}(\alpha)c_{\mu}(\alpha}}}\alpha) c_{\mu}(\underbracket{\alpha)c_{\mu}(\beta)c_{\mu}(\beta})\rangle_{V}=\Lambda^{2}(\mu,\alpha)\Lambda(\mu,\alpha_{0})\delta_{\alpha\beta}.
\end{equation}
Hence we get
\begin{eqnarray}
&&\sum_{\mu}\sum_{\alpha\beta}\langle c_{\mu}(\overbracket{\alpha_{0})c_{\mu}(\alpha_{0}}) c_{\mu}(\lefteqn{\overbracket{\phantom{\alpha)c_{\mu}(\alpha)c_{\mu}(\alpha}}}\alpha) c_{\mu}(\underbracket{\alpha)c_{\mu}(\beta)c_{\mu}(\beta})\rangle_{V}(\hat{H}^{\prime}_{0})_{\alpha\alpha}(\hat{H}^{\prime}_{0})_{\beta\beta}
=\sum_{\mu}(\sum_{\alpha}\Lambda^{2}(\mu,\alpha)(\hat{H}^{\prime}_{0})^{2}_{\alpha\alpha})\Lambda(\mu,\alpha_{0})\notag\\
&&\approx [(\hat{H}^{\prime}_{0})^{2}_{\alpha\alpha}]_{\mu}\sum_{\mu}\sum_{\alpha}\Lambda^{2}(\mu,\alpha)\Lambda(\mu,\alpha_{0})=\frac{\omega}{2\pi\Gamma}[(\hat{H}^{\prime}_{0})^{2}_{\alpha\alpha}]_{\mu}.\label{a}
\end{eqnarray}
Similarly we have 
\begin{equation}
\sum_{\mu}\sum_{\alpha\beta}\langle c_{\mu}(\overbracket{\alpha_{0})c_{\mu}(\alpha_{0}}) c_{\mu}(\overbracket{\alpha)c_{\mu}(\underbracket{\alpha)c_{\mu}(\beta})c_{\mu}(\beta})\rangle_{V}(\hat{H}^{\prime}_{0})_{\alpha\alpha}(\hat{H}^{\prime}_{0})_{\beta\beta}\approx
\frac{\omega}{2\pi\Gamma}[(\hat{H}^{\prime}_{0})^{2}_{\alpha\alpha}]_{\mu}.\label{b}
\end{equation}
Other six-point correlation term is 
\begin{eqnarray}
&&\sum_{\mu}\sum_{\alpha\beta}\langle c_{\mu}(\overbracket{\alpha_{0})c_{\mu}(\alpha_{0}}) c_{\mu}(\overbracket{\alpha)c_{\mu}(\alpha})c_{\mu}(\overbracket{\beta)c_{\mu}(\beta})\rangle_{V}(\hat{H}^{\prime}_{0})_{\alpha\alpha}(\hat{H}^{\prime}_{0})_{\beta\beta}=\sum_{\mu}\sum_{\alpha\beta}
\Lambda(\mu,\alpha)\Lambda(\mu,\beta)\Lambda(\mu,\alpha_{0})
(\hat{H}^{\prime}_{0})_{\alpha\alpha}(\hat{H}^{\prime}_{0})_{\beta\beta}\notag\\
&&\approx [(\hat{H}^{\prime}_{0})_{\alpha\alpha}]^{2}_{\mu}\sum_{\mu}\sum_{\alpha\beta}\Lambda(\mu,\alpha)\Lambda(\mu,\beta)\Lambda(\mu,\alpha_{0})=[(\hat{H}^{\prime}_{0})_{\alpha\alpha}]^{2}_{\mu}.\label{c}
\end{eqnarray}
Combining Eqs. (\ref{a}), (\ref{b}), and (\ref{c}) we obtain Eq. (\ref{a1}). Now, let's consider the corrections. We have

\begin{equation}
\langle c_{\mu}(\underbracket{\alpha_{0}) c_{\mu}(\overbracket{\alpha_{0})c_{\mu}(\alpha})c_{\mu}(\overbracket{\alpha)c_{\mu}(\beta})c_{\mu}(\beta})\rangle_{V}=\Lambda(\mu,\beta)\delta_{\alpha_{0},\beta}\Lambda(\mu,\alpha_{0})\delta_{\alpha,\alpha_{0}}
\Lambda(\mu,\alpha)\delta_{\alpha\beta}.
\end{equation}
Therefore, we obtain
\begin{equation}
\sum_{\mu}\sum_{\alpha\beta}\langle c_{\mu}(\underbracket{\alpha_{0}) c_{\mu}(\overbracket{\alpha_{0})c_{\mu}(\alpha})c_{\mu}(\overbracket{\alpha)c_{\mu}(\beta})c_{\mu}(\beta})\rangle_{V}
(\hat{H}^{\prime}_{0})_{\alpha\alpha}(\hat{H}^{\prime}_{0})_{\beta\beta}=(\hat{H}^{\prime}_{0})^{2}_{\alpha_{0}\alpha_{0}}\sum_{\mu}\Lambda^{3}(\mu,\alpha_{0}).
\end{equation}
Such a term gives correction of order of $(\omega/\Gamma)^{2}$. In fact there are in total $8$ terms which give correction of such order. 

Consider now the term
\begin{equation}
\langle c_{\mu}(\overbracket{\alpha_{0})c_{\mu}(\underbracket{\alpha_{0})c_{\mu}(\alpha})c_{\mu}(\alpha})c_{\mu}(\overbracket{\beta)c_{\mu}(\beta})\rangle_{V}=\Lambda^{2}(\mu,\alpha)\delta_{\alpha,\alpha_{0}}\Lambda(\mu,\beta).
\end{equation}
Hence we get
\begin{eqnarray}
&&\sum_{\mu}\sum_{\alpha\beta}\langle c_{\mu}(\overbracket{\alpha_{0})c_{\mu}(\underbracket{\alpha_{0})c_{\mu}(\alpha})c_{\mu}(\alpha})c_{\mu}(\overbracket{\beta)c_{\mu}(\beta})\rangle_{V}(\hat{H}^{\prime}_{0})_{\alpha\alpha}
(\hat{H}^{\prime}_{0})_{\beta\beta}=\sum_{\mu}\sum_{\beta}\Lambda(\mu,\beta)(\hat{H}^{\prime}_{0})_{\beta\beta}\Lambda^{2}(\mu,\alpha_{0})(\hat{H}^{\prime}_{0})_{\alpha_{0}\alpha_{0}}\notag\\
&&\approx [(\hat{H}^{\prime}_{0})_{\alpha\alpha}]_{\mu}(\hat{H}^{\prime}_{0})_{\alpha_{0}\alpha_{0}}\sum_{\mu}\sum_{\beta}\Lambda(\mu,\beta)\Lambda^{2}(\mu,\alpha_{0})=\frac{\omega}{2\pi\Gamma}[(\hat{H}^{\prime}_{0})_{\alpha\alpha}]_{\mu}(\hat{H}^{\prime}_{0})_{\alpha_{0}\alpha_{0}}.
\end{eqnarray}
There are in total $4$ terms with the same contribution. Note that for spin chain that we consider the microcanonical average of the spin observable is zero and thus these terms can be neglected.

\end{section}

\begin{section}{Spin-Spin Correlations}
\begin{figure}
\includegraphics[width=0.5\textwidth]{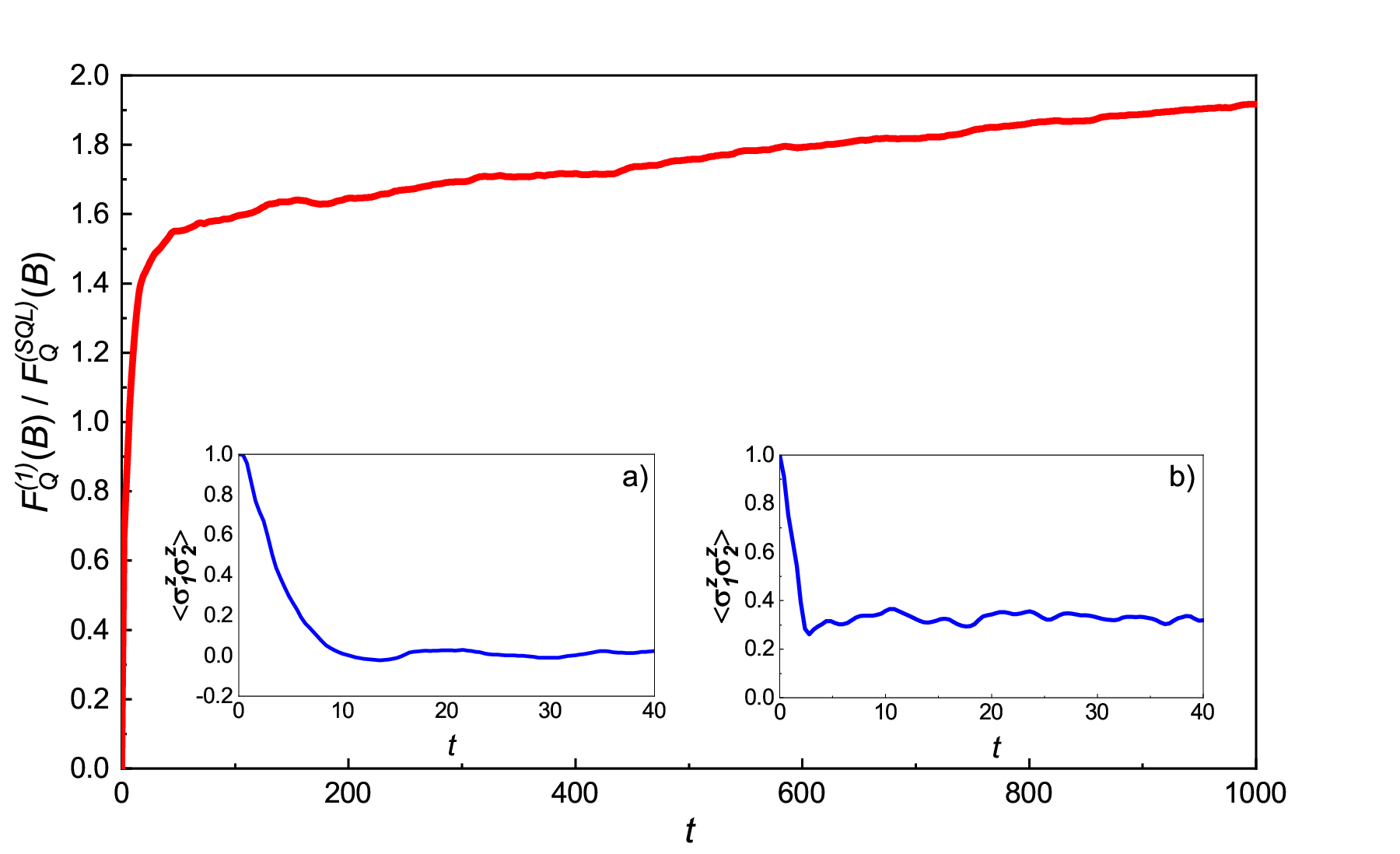}
\caption{Exact time evolution of the ratio $F_Q^{(1)}/F_Q^{(\rm SQL)}$ for a spin chain consisting of two system spins. The initial state is $|\Psi_{0}\rangle=\left|\uparrow \uparrow\right\rangle_S \left|\uparrow\downarrow \uparrow ... \right\rangle_B$ for $N=15$.  (Case 1) The system spins interact with different bath spins ($r_{1}=5$ and $r_{2}=8$) with QFI $F_Q^{(\rm SQL)}$. (Case 2) Both systems spins interact with the same bath spin ($r_{1}=5$) with corresponding QFI $F_Q^{(1)}$. Time evolution of the correlation $\langle\sigma^{z}_{1}\sigma^{z}_{2}\rangle$ for cases 1 (inset (a)) and 2 (inset (b)).} 
\label{fig6}
\end{figure}
Here we extend the discussion by consider a spin system Hamiltonian consisting of two spins
\begin{equation}
\hat{H}_{\rm S}=B(\sigma^{z}_{1}+\sigma^{z}_{2})
\end{equation}
and spin-bath interaction
\begin{eqnarray}
\hat{H}_{\rm SB}=J_z^{(\rm SB)}\sigma^{z}_{1}\sigma^{z}_{r_{1}}+J_x^{(\rm SB)}(\sigma^{+}_{1}\sigma^{-}_{r_{1}}+\sigma^{-}_{1}\sigma^{+}_{r_{1}})
+J_z^{(\rm SB)}\sigma^{z}_{2}\sigma^{z}_{r_{2}}+J_x^{(\rm SB)}(\sigma^{+}_{2}\sigma^{-}_{r_{2}}+\sigma^{-}_{2}\sigma^{+}_{r_{2}}),
\end{eqnarray}
where $r_{k}$ ($k=1,2$) denotes the position of the bath spins. For two spins coupled to different bath spins $r_{1}\neq r_{2}$ we find that no correlation is created between the system spins in the sense that $(\sigma^{z}_{1}\sigma^{z}_{2})_{\rm mc}=(\sigma^{z}_{1})_{\rm mc}(\sigma^{z}_{2})_{\rm mc}\approx 0$, see Fig. (\ref{fig6}) (inset (a)). In that case, the QFI is twice the QFI for a single system spin, $F^{(\rm SQL)}_{Q}(B)=2 F_{Q}(B)$, which corresponds to the standard quantum limit (SQL). Let us now consider $r_{1}=r_{2}$, namely, the two system spins are coupled to the single bath spin. Then the spin-bath interaction creates a correlation between the two system spins in a sense that $(\sigma^{z}_{1}\sigma^{z}_{2})_{\rm mc}\neq 0$, see Fig. (\ref{fig6}) (inset (b)). Using Eq. (\ref{main}) with $\hat{H}^{\prime}_{0}=\sigma^{z}_{1}+\sigma^{z}_{2}$ we obtain
\begin{eqnarray}
F^{(1)}_{Q}(B)=F^{(\rm SQL)}_{Q}(B)+4t^{2}(\sigma^{z}_{1}\sigma^{z}_{2})_{\rm mc}
\left\{\frac{2}{\pi D(E_{0})\Gamma}
+\frac{1}{(\Gamma t)^{2}}
(e^{-2\Gamma t}-1+2\Gamma t)\right\}.
\end{eqnarray}
As long as $(\sigma^{z}_{1}\sigma^{z}_{2})_{\rm mc}>0$ we have $F^{(1)}_{Q}(B)>F^{(\rm SQL)}_{Q}(B)$ and thus one can overcome the SQL. We plot in Fig. \ref{fig6} the exact time evolution of the ratio $F^{(1)}_{Q}/F^{(\rm SQL)}_{Q}$. We see that the positive quantum correlation between the system spins leads to enhancement of the QFI compare with the $F^{(\rm SQL)}_{Q}(B)$.

\end{section}

\begin{section}{Defining a local observable in RMT}

In the main we have analysed the QFI for a \emph{local} observable of the system interacting with a bath. In this section we will see that such an approach can be formalised within our RMT approach, and indeed gives way to a crucial condition - observable sparsity - of the application of RMT. We will see that an additional condition is required on the system and bath parts of the total system in order for RMT to apply to such local observables, namely, that the system energy is much smaller than that of the bath.

We begin by separating the system into system and bath components via $\hat{H}_0 = \hat{H}_S(\lambda) \otimes \mathbf{1}_B + \mathbf{1}_{S} \otimes \hat{H}_B$, with $\mathbf{1}_{S(B)}$ the identity on the system (bath) Hilbert space. Crucially, here the system part of the Hamiltonian is assumed to depend on some parameter $\lambda$. The eigenstates of $\hat{H}_0$ are then
\begin{align}
    |\phi_\alpha\rangle = |s(\alpha)\rangle_S \otimes |\phi_{\alpha_B(\alpha)}^{(B)}\rangle_B,
\end{align}
with energies 
\begin{align}
    E_{\alpha} & = {}_S\langle s(\alpha)| {}_B \langle \phi_{\alpha_B(\alpha)}^{(B)} | \hat{H}_S(\lambda) + \hat{H}_B |s(\alpha)\rangle_S \otimes |\phi_{\alpha_B(\alpha)}^{(B)}\rangle_B \nonumber \\&
    = \epsilon_{s(\alpha)}(\lambda) + E^{(B)}_{\alpha_B(\alpha)},
\end{align}
where we have denoted eigenenergies of the system and bath Hamiltonians by $\epsilon_{s(\alpha)}(\lambda)$ and $E^{(B)}_{\alpha_B(\alpha)}$ respectively.

Relevant observables in our approach act on the system Hilbert space as $\hat{O} = \hat{O}_S \otimes \mathbf{1}_B$, which have matrix elements
\begin{align}
    O_{\alpha\beta} &= {}_S\langle s(\alpha)| {}_B \langle \phi_{\alpha_B(\alpha)}^{(B)} | \hat{O} |s(\beta)\rangle_S \otimes |\phi_{\alpha_B(\beta)}^{(B)}\rangle_B \nonumber \\&
    = (O_S)_{s(\alpha) s(\beta)} \delta_{\alpha_B(\alpha) \alpha_B(\beta)},
\end{align}
where $(O_S)_{s(\alpha) s(\beta)} =  {}_S\langle s(\alpha)| \hat{O}_S |s(\beta)\rangle_S $. We see here that the local observable $\hat{O}$ is guaranteed to be sparse if the dimension of the system Hilbert space $d_S$ is much lower than that of the bath, $d_B$, as there are a maximum of $d_S (d_S - 1)$ independent off diagonal matrix elements of $O_{\alpha\beta}|_{\alpha\neq\beta}$, corresponding to the possible system state transitions, plus $d_S$ possible diagonal matrix elements.

These possible transitions that the local operator $\hat{O}$ may induce must obey
\begin{align}
    E_\alpha - E_\beta =  \epsilon_{s(\alpha)}(\lambda) -  \epsilon_{s(\beta)}(\lambda) := \Delta_{\alpha\beta}^{(S)}(\lambda),
\end{align}
and more generally, we have,
\begin{align}
    E_\alpha - E_\beta = \Delta_{\alpha\beta}^{(S)}(\lambda) + E^{(B)}_{\alpha_B(\alpha)} - E^{(B)}_{\alpha_B(\beta)}.
\end{align}
For the random matrix model, we have $E_\alpha = \alpha \omega$, so we require for the RMT to hold that $E_\alpha - E_\beta \approx (\alpha - \beta)\omega$ can be approximated by an equidistant spacing of energies that does not depend on $\lambda$. This is understood to hold if $\hat{H}_B$ is itself a non-integrable Hamiltonian, and if 
$\Delta_{\alpha\beta}^{(S)}(\lambda) \ll E^{(B)}_{\alpha_B(\alpha)} - E^{(B)}_{\alpha_B(\beta)}$, indicating that the possible transitions induced by the local observable are negligible in energy in comparison to the bath energy for the state $|\phi_{\alpha}\rangle$.
\end{section}

\end{document}